\documentclass[11pt]{article}
\pdfoutput=1

\usepackage[usenames]{color}
\usepackage{fix-cm}
\usepackage{booktabs}

\newcommand{\Feyn}[1]{#1\kern-0.45em/}

\usepackage[english]{babel}
\usepackage{amsmath,amssymb,amsbsy,amstext, amsthm}
\usepackage{graphicx}
\usepackage{amsfonts}
\usepackage{amssymb}
\usepackage{subfigure}
\usepackage{color}
\usepackage{float}
\usepackage[small]{caption}

\numberwithin{equation}{section}

\def\beq{\begin{equation}}
\def\eeq{\end{equation}}

\def\bea{\begin{eqnarray}}
\def\eea{\end{eqnarray}}

\def\d{{\rm d}}

\def\beq{\begin{equation}}
\def\eeq{\end{equation}}
\def\bea{\begin{eqnarray}}
\def\eea{\end{eqnarray}}

\def\d{{\rm d}}

\def\es{\epsilon_s}

\DeclareRobustCommand{\SkipTocEntry}[4]{}

\begin{document}

\begin{titlepage}

\setcounter{page}{1} \baselineskip=15.5pt \thispagestyle{empty}

\bigskip\
\vspace{2cm}
\begin{center}
{\fontsize{16}{30}\selectfont  \bf  Scale-Invariance and the Strong Coupling Problem}
\end{center}

\vspace{0.5cm}
\begin{center}
{\fontsize{13}{30}\selectfont  Daniel Baumann$^\clubsuit$, Leonardo Senatore$^{\diamondsuit,\spadesuit}$, and Matias Zaldarriaga$^\clubsuit$}
\end{center}

\begin{center}
\vskip 8pt
\textsl{$^\clubsuit$ School of Natural Sciences,
 Institute for Advanced Study,
Princeton, NJ 08540}

\vskip 7pt
\textsl{$^\diamondsuit$ Stanford Institute for Theoretical Physics, \\ Stanford University, Stanford, CA 94305}

\vskip 7pt
\textsl{$^\spadesuit$ Kavli Institute for Particle Astrophysics and Cosmology, \\ Stanford University, 
Stanford, CA 94305}

\end{center} 

\vspace{1.2cm}
\hrule \vspace{0.3cm}
{ \noindent \textbf{Abstract} \\[0.1cm]
\noindent
The effective theory of adiabatic fluctuations around arbitrary Friedmann-Robertson-Walker backgrounds---both expanding and contracting---allows for more than one way to obtain scale-invariant two-point correlations. However, as we show in this paper, it is challenging to produce scale-invariant fluctuations that are
weakly coupled over the range of wavelengths accessible to cosmological observations.
In particular, requiring the background to be a dynamical attractor, the curvature fluctuations are scale-invariant and weakly coupled for at least 10~$e$-folds only if the background is close to de Sitter space. In this case, the time-translation invariance of the background guarantees time-independent $n$-point functions. For non-attractor solutions, any predictions depend on assumptions about the evolution of the background even when the perturbations are outside of the horizon. For the simplest such scenario we identify the regions of the parameter space that avoid both classical and quantum mechanical strong coupling problems. Finally, we present extensions of our results to backgrounds in which higher-derivative terms play a significant role.}
 \vspace{0.3cm}
 \hrule

\vspace{0.6cm}

\end{titlepage}

\tableofcontents

\newpage

\section{Introduction}

A key feature of the primordial seed fluctuations observed in the cosmic microwave background (CMB) and the large-scale structure (LSS) is their {\it scale-invariance}~\cite{WMAP,SDSS}.
Furthermore, the observed near-Gaussianity of the data requires the fluctuations to be {\it weakly coupled}\,\footnote{`Weak coupling' implies that the action for the fluctuations has a well-defined expansion in which the leading terms of order $n+1$ in the fluctuations are smaller than the terms of order $n$. We shall be more precise about this in \S\ref{sec:strong}.} over a large range of scales. In this paper we show that if the universe was dominated by a single degree of freedom at the time when the fluctuations were created, the above two facts together strongly constrain the background spacetime at that time. 

\vskip 4pt
Our basic point is very simple: 
the scale-invariance of two-point correlations does not guarantee scale-invariance of interactions.
Observations require scale-invariance of the power spectrum of curvature perturbations over at least $\Delta N \sim 10$ $e$-folds, from CMB scales ($\sim 10^4$\,Mpc) to galactic scales ($\sim 1$\,Mpc).
These fluctuations are created while the conformal time $\tau$ evolves by a factor of $e^{-\Delta N}$. 
In quasi-de Sitter backgrounds, this exponential change in time appears only in the time-evolution of the scale factor $a(\tau) = -(H \tau)^{-1}$, which is not a physical observable. All observable couplings during inflation are nearly time-independent. This is a consequence of the time-translation invariance of the background. The scale-invariance of the fluctuations produced by inflation therefore applies to all $n$-point functions.
In contrast, in non-de Sitter backgrounds with a single fluctuating degree of freedom\footnote{All of our constraints can be evaded in {\it multi-field} non-de Sitter cosmologies such as~\cite{Creminelli:2007aq, Buchbinder:2007ad, Lehners:2007wc}.} (e.g.~\cite{PaulJustin, Justin2}), it is hard to achieve both scale-invariant two-point correlations and time-independent interactions  simultaneously. 
This is the case because the $n$-point functions of adiabatic (or `single-clock') fluctuations around Friedmann-Robertson-Walker (FRW) backgrounds aren't independent, but are related by the symmetries of the background~\cite{Creminelli:2006xe, Cheung}: 
time translations are spontaneously broken in FRW backgrounds  and the symmetry is {\it non-linearly} realized by  a Goldstone boson. 
This non-linear realization of time translations forces unavoidable relations between the quadratic terms and the higher-order terms in the Lagrangian. The interactions are particularly constrained when  the Goldstone boson of time translations is  the only relevant degree of freedom.  It is precisely these constraints that will allow us to make statements about the interaction Lagrangian in any cosmology that produces a scale-invariant two-point function.
We find that any time-dependence that is not in the scale factor typically leads to an exponential growth of interactions, so that one has to be worried that the fluctuations become strongly coupled at some point during the evolution.
At this point perturbative control of the theory would be lost.

\vskip 4pt
In Section~\ref{sec:two} we review the conditions for scale-invariant two-point correlations in the effective theory of adiabatic fluctuations. 
At lowest order in derivatives, we identify two classes of backgrounds that allow for scale-invariant two-point functions for the primordial curvature perturbation $\zeta$.
These two possibilities are distinguished by whether $\zeta$ is constant on superhorizon scales ({\sf Case I}) or grows in time by an exponentially large amount ({\sf Case II}). 
These two possibilities correspond to the background being a dynamical attractor or not.
Within these two classes of solutions there are important subclasses depending on whether the background is expanding or contracting, has a strongly time-dependent equation of state or a strongly time-dependent speed of sound.
In Section~\ref{sec:strong} we explain that higher-order interactions of $\zeta$ and the associated higher-point correlations severely limit the possibilities.
We present the leading cubic interactions, identify the time-dependent `coupling constants', and discuss when they lead to a strong coupling problem.
In Section~\ref{sec:attractor} we study the attractor solutions in detail. For the case of a constant speed of sound we find {\it exact} solutions which allow us to study FRW backgrounds in full generality. 
We show that only near-de Sitter backgrounds, with intrinsically small time-variations of all quantities, avoid the strong coupling problem.
We explain why our conclusions are not affected significantly when the speed of sound is allowed to vary in time.
We present our conclusions in Section~\ref{sec:conclusions}.

\vskip 4pt
Four appendices contain further technical details:

In Appendix~\ref{sec:A2} we derive the quadratic action for the curvature perturbation $\zeta$ in the framework of the effective theory of adiabatic fluctuations~\cite{Creminelli:2006xe, Cheung}. We show that, at leading order in a derivative expansion, the dynamics is characterized by two free functions: the Hubble rate $H(t)$ and the sound speed $c_s(t)$.
In the limit in which $M_{\rm pl}^2 \dot H$ becomes smaller than other parameters in the Lagrangian, higher-derivative terms can be the leading effects. We discuss this limit in Appendix~\ref{sec:new}. 
In Appendix~\ref{sec:nonA} we turn our attention to the non-attractor cases.
These theories are significantly less predictive and require a variety of additional assumptions about the evolution both before and after the scale-invariant modes are produced.
We will show that even granting the most optimistic such assumptions 
the parameter space is significantly constrained by the weak-coupling requirement.
Finally, in Appendix~\ref{sec:bispectra} we compute the full bispectra for the non-attractor cases of theories with non-trivial speed of sound, following earlier work by Khoury and Piazza~\cite{Justin}.

\section{Scale-Invariance of the Two-Point Function}
\label{sec:two}

Although originally introduced to study the possibility of violating the null energy condition~\cite{Creminelli:2006xe} or to describe adiabatic fluctuations around inflationary backgrounds~\cite{Cheung}, the effective theory of adiabatic fluctuations applies to arbitrary FRW backgrounds. 
It is therefore a promising starting point to discuss the properties of fluctuations around a wide class of background spacetimes, both expanding and contracting.
Using this framework we show in Appendix~\ref{sec:A2} that the quadratic action for curvature perturbations $\zeta$ (at leading order in a derivative expansion of metric perturbations in comoving gauge) depends only on two free functions of time: the scale factor $a(t)$ and the speed of sound $c_s(t)$. 
In addition, we have the following derived quantities
\beq
\label{equ:SRpara}
\epsilon = - \frac{\dot H}{H^2}\ , \qquad \eta = \frac{\dot \epsilon}{H \epsilon} \ , \qquad \epsilon_s = \frac{\dot c_s}{H c_s}\ ,
\eeq
where dots are derivatives with respect to time $t$ and $H \equiv \partial_t \ln a$ is the Hubble parameter. During 
inflation the parameters in (\ref{equ:SRpara}) are all much less than unity and the speed of sound is nearly constant, but here were want to consider a broader spectrum of possibilities.

Our starting point is the quadratic action for curvature fluctuations (see Appendix~\ref{sec:A2})
\beq
S_{2} =  M_{\rm pl}^2 \int \d t\, \d^3 x \, \frac{a^3 \epsilon}{c_s^2} \left[ \dot \zeta^2  - \frac{c_s^2}{a^2} (\partial_i \zeta)^2 \right] \, .
\eeq
Rescaling time, 
\beq
\d y = c_s \d \tau = \frac{c_s}{a} \d t\ ,
\eeq 
this can be written as
\beq
\label{equ:action2}
\fbox{$\displaystyle S_{2} = M_{\rm pl}^2 \int \d^4 x \ q^2 \left[  (\zeta')^2 -  (\partial_i \zeta)^2 \right] $}\, ,
\eeq
where 
\beq
\fbox{$\displaystyle q^2 \equiv  \frac{a^2 \epsilon}{c_s} $}\ .
\eeq
The prime in (\ref{equ:action2}) indicates a derivative with respect to $y$ and $\d^4 x \equiv \d y \, \d^3 x$.
The time variable $y$ is convenient because a mode of comoving wavenumber $k$ crosses the (sound) horizon when $|k y_\star| = 1$. 
We are interested in the regime where modes cross the shrinking horizon $|y|$, so that we concentrate on negative $y$.
The behavior of $\zeta_k(y)$ changes from oscillating modes to monotonically growing and decaying modes at $y_\star$.
The prefactor $q^2$ in (\ref{equ:action2}) has to be positive: a negative $q^2$ implies a wrong-sign kinetic term for $\zeta$ and hence a ghost instability.
Notice that a negative $q^2$ implies a violation of the null energy condition ($\dot H>0$). This can be achieved only using higher-derivative terms in the effective action~\cite{Creminelli:2006xe}.

The time-dependence of the prefactor $q(y)$ determines whether the fluctuations have scale-invariant two-point correlations.
To see this, let us define the canonically-normalized variable $v \equiv \sqrt{2}\, q \zeta$. 
After a few integrations by parts the action becomes
\beq
\label{equ:action3}
S_{2} = \frac{M_{\rm pl}^2}{2} \int \d^4 x  \,  \left[  (v')^2 -  (\partial_i v)^2 + \frac{q''}{q} v^2 \right] \ .
\eeq
This corresponds to the following equation of motion in Fourier space
\beq
\label{eom}
v_k'' + \left[ k^2 - \frac{q''}{q}\right] v_k = 0\, ,
\eeq
where
\beq
\label{equ:qq}
\frac{q''}{q} = \frac{n(n-1)}{y^2} \qquad {\rm if} \qquad q \propto y^n\ .
\eeq
For the power law ansatz (\ref{equ:qq}), the differential equation (\ref{eom}) has an exact solution in terms of Hankel functions of the first and second kind:
\beq
v_k(y) = \sqrt{-y}\, \bigl[ c_1H_\nu^{(1)}(-ky) + c_2 H_\nu^{(2)}(-ky) \bigr]\ , \qquad {\rm where} \quad \nu^2 - \frac{1}{4} \equiv n(n-1)\ . 
\eeq
Demanding that the solution reduces to the standard Minkowski limit on small scales (or at early times), 
\beq
\label{lim1}
\lim_{ky \to -\infty} v_k(y) =\frac{e^{-iky}}{\sqrt{2 k}} \ ,
\eeq
fixes $c_2 = 0$ and $c_1 = \sqrt{\frac{\pi}{2}}$. 
The power spectrum on superhorizon scales then is
\beq
\label{lim2}
\lim_{ky \to 0} k^3 |v_k|^2 \sim  \frac{1}{y^2}  (-ky)^{3-2 \nu} \ ,
\eeq
where we used $\lim_{ky \to 0} H_\nu^{(1)} \sim (-k y)^{-\nu}$. This describes scale-invariant fluctuations if and only if
\beq
\nu = \frac{3}{2} \qquad \Rightarrow \qquad \nu^2 - \frac{1}{4} = 2 = n(n-1) \qquad \Rightarrow \qquad n = \left\{ \, -1\, , \, 2 \, \right\}\ .
\eeq
We hence found two distinct cases for which the background $q(y)$ leads to a scale-invariant spectrum of fluctuations~\cite{Justin}:\,\footnote{These solutions may also be found without assuming $q''/q$ to be a power law, by requiring invariance of the action~(\ref{equ:action2}) under the rescaling $y \to \lambda y$, $x \to \lambda x$ and $\zeta\rightarrow \lambda^n\zeta$, with $n$ some number. If the initial state does not break this symmetry---as is the case for the Bunch-Davies vacuum---then the two-point functions of fields with scaling dimension 0 are scale-invariant.}
\begin{align}
&{\sf Case\ I}   \qquad  \ \, \fbox{$\displaystyle q = q_i \left(\frac{y}{y_i}\right)^{-1} $} \ , \label{Case1}\\
&{\sf Case\ II}  \qquad \, \fbox{$\displaystyle q = q_i \left(\frac{y}{y_i}\right)^{2} \hskip 7pt $} \ . \label{Case2}
\end{align}
On superhorizon scales the growing mode of $\zeta$ is constant in {\sf Case I} and time-dependent in {\sf Case~II}:
\beq
\lim_{ky \to 0} \zeta = \lim_{ky \to 0} \frac{v}{q} = \zeta_i \left\{ 
\begin{array}{l}
\ \ \,  1 \qquad \ \ \ \ \  \, {\sf Case\ I}\\
 \bigl(\frac{y}{y_i}\bigr)^{-3}  \qquad \ {\sf Case\ II}
\end{array}
\right.
\eeq
This difference in the time-dependence of $\zeta$ reflects the (in)stability of the background.
In {\sf Case I}, $\zeta$ approaches a constant on superhorizon scales and the background is a stable attractor solution;
in the long-wavelength limit, $k \to 0$, classical solutions for $\zeta$ are interpreted as homogeneous perturbations to the background scale factor: $\d s^2 = a^2(\tau) (- \d \tau^2 + e^{2\zeta(\tau,{\bf x})} \d {\bf x}^2)$.
In contrast, the exponentially large growth of $\zeta$ in {\sf Case II} indicates that the background is unstable and not an attractor.
This instability implies that {\sf Case II} is not predictive unless additional assumptions are made about the evolution before and after the scale-invariant modes exit the horizon (see Appendix~\ref{sec:nonA}).

Several examples of backgrounds satisfying (\ref{Case1}) and (\ref{Case2}) are known:
for constant speed of sound $c_s = const.$ the condition for scale-invariance, 
\beq
\label{equ:condX}
q^2 = \frac{a^2\epsilon}{c_s} = \frac{a^2}{c_s} \left[ 2 - \frac{a'' a}{(a')^2} \right] \propto \tau^{-2}\ ,
\eeq
 implies that either the scale factor $a$ or the equation of state $\epsilon$ (or a combination of both) vary rapidly with time. The case $a \propto - \tau^{-1}$ and $\epsilon \approx const.$ of course corresponds to slow-roll inflation~\cite{TASI}, while $\epsilon \propto \tau^{-2}$ and $a \approx const.$ corresponds to adiabatic ekpyrotic contraction (or expansion) with time-varying equation of state \cite{PaulJustin, Justin2}. In Section~\ref{sec:attractor} we will find an {\it exact} solution to (\ref{equ:condX}) which to our knowledge is new. This solution contains inflation and adiabatic ekpyrosis as special cases.
Allowing for a non-trivial speed of sound, Khoury and Piazza~\cite{Justin} discussed the production of scale-invariant fluctuations in non-attractor backgrounds.
Because of the many extra assumptions and model-building requirements, we find the non-attractor solutions much less compelling than the attractor solutions. 
In the main text we will therefore focus on the attractors cases, leaving a detailed treatment of the non-attractor cases to Appendix~\ref{sec:nonA}.

\section{The Strong Coupling Problem}
\label{sec:strong}

The main point of our paper is the following: 
although there are many different FRW backgrounds -- both expanding and contracting -- that lead to scale-invariant two-point correlations (see Section~\ref{sec:attractor} and Appendix~\ref{sec:nonA}), in most cases instabilities are revealed when looking at higher-order correlations.
In other words, scale-invariance of the two-point function does not guarantee scale-invariance of the interactions. This is the case because the non-linear realization of time-diffeormorphisms forces non-trivial connections between the quadratic and the cubic Lagrangian.

\vskip 4pt
The cubic Lagrangian contains among others the following terms~\cite{Chen} (see Appendix~\ref{sec:bispectra})
\beq
\label{equ:cubicX}
{\cal L}_3 \ \subset\  \frac{a^3 \epsilon}{c_s^2} \left[ \frac{1}{c_s^2} \big(\epsilon - 3(1-c_s^2) \big) \, \zeta \dot \zeta^2 + \big(\epsilon-2 \epsilon_s +1-c_s^2 \big)\, \zeta \frac{(\partial \zeta)^2}{a^2} + \frac{1}{2c_s^2} \big(\dot \eta + H \eta - 2 H \epsilon_s \big) \, \zeta^2 \dot \zeta \right]\, ,
\eeq
where $S_3 = \int \d t \, \d^3 x \, {\cal L}_3$.
As a simple measure of our strong coupling concern we use the ratio of cubic to quadratic Lagrangian:
\beq
\label{equ:X}
\fbox{$\displaystyle X \equiv \frac{{\cal L}_{3}}{{\cal L}_{2}} \ \sim\ {\cal O}(\{1, \epsilon, \eta, \epsilon_s \}) \, \frac{\zeta}{c_s^2} $} \ \ll \ 1 \ .
\eeq
When this ratio (or more generally the ratio ${\cal L}_{\rm int}/{\cal L}_2$) becomes larger than one, the theory is strongly coupled. There are two different ways in which having ${\cal L}_{\rm int}/{\cal L}_2$ larger than one is dangerous. The first is when this condition happens {\it at} horizon crossing. In this case the theory is strongly coupled at the quantum level, in the sense that quantum loop corrections aren't suppressed~\cite{Cheung,Sarah}. This is the regime which is usually referred to as strong coupling. In the absence of a UV-completion or a proof of strongly coupled unitarization, unitarity is lost and the model is not under theoretical control. For attractor solutions ({\sf Case I}), this is the only condition that we will require.

For non-attractor solutions ({\sf Case II}), the condition (\ref{equ:X}) can also be violated {\it after} horizon crossing. In this case, ${\cal L}_{\rm int} > {\cal L}_2$ does not represent a violation of unitarity or an actual quantum mechanical strong coupling. It simply signals the presence of classical non-linearities. In principle, these can be treated by solving the non-linear equations for the modes. Hence, strictly speaking, the model is still under theoretical control. However, the predictions that are made by studying only the quadratic Lagrangian, including the prediction of scale invariance,  can't be trusted anymore. Moreover, if such a regime happens for the observable modes, then the fluctuations can be strongly non-Gaussian and in conflict with observations.\footnote{Recall that to be consistent with present microwave background and large-scale structure observations the background has to allow for at least of order $\Delta N \approx10$~$e$-folds of scale-invariant and weakly-coupled fluctuations.} We will need to use this classical strong coupling condition only when studying the non-attractor solutions ({\sf Case~II}) in  Appendix~\ref{sec:nonA}.

\vskip 4pt 
 In {\sf Case I}, $\zeta$ freezes at horizon-crossing, so that its magnitude at that time is fixed by observations, $\zeta_\star \sim 10^{-5}$. Similarly, the speed of sound at horizon-crossing can't be too small,  $c_{s,\star} > \zeta_\star^{1/2} \sim 10^{-3}$, to avoid strong coupling from interactions proportional to $c_s^{-2}$\,(\footnote{Observational constraints on non-Gaussianity~\cite{SenatoreSmithWMAP5} of course imply an even stronger constraint on the minimal speed of sound, $c_{s,\star} \gtrsim 0.01$.}). The constraint $X\ll1$ then becomes a constraint on the size of the `couplings' $\epsilon$, $\eta$ and $\epsilon_s$~\cite{SenatoreSmithWMAP5}. 
 If these couplings are time-dependent one has to worry that the theory becomes strongly coupled at some time in the regime of cosmological interest.
 We discuss this in detail in Section~\ref{sec:attractor}.
 
 \vskip 4pt
In {\sf Case II}, the analysis is significantly complicated by the fact that $\zeta$ evolves outside of the horizon. Its value at horizon-crossing, $\zeta_\star$, can be exponentially smaller than its value at the end of the scaling regime, $\zeta_{\rm end}$.
 Whether the fluctuations are weakly or strongly coupled at horizon-crossing then depends on the time-evolution of the speed of sound. Furthermore, the superhorizon evolution of $\zeta$ can lead to additional classical non-linearities.
In Appendix~\ref{sec:nonA} we discuss the combined constraints from non-linearities at horizon crossing and the subsequent classical evolution.

\section{Scale-Invariance for Dynamical Attractors}
\label{sec:attractor}

Only in {\sf Case I} is the background a dynamical attractor.
This means that curvature fluctuations freeze at horizon crossing and the predictions for observations are independent of the subsequent evolution.
We will first consider the limit of constant sound speed, $c_s \approx const.$, and then explain why a time-dependent sound speed does not qualitatively change our conclusions.

\subsection{Emden-Fowler Equation}

As we have shown above, 
theories with constant sound speed produce scale-invariant fluctuations if the scale factor $a(\tau)$ satisfies the following non-linear second-order differential equation,
\beq
\label{equ:a}
q^2 = \frac{a^2 \epsilon}{c_s} = \frac{a^2}{c_s} \left[2 - \frac{a'' a}{(a')^2} \right] = q_i^2\, \frac{\tau_i^2}{\tau^2}
\ ,
\eeq
where $'$ denotes derivatives with respect to conformal time $\tau$. 
We assume that (\ref{equ:a}) is valid in a finite time-interval between $\tau_i$ and $\tau_{\rm end}$. 
During this `scaling regime' the background is consistent with scale-invariant two-point correlations.
It is convenient to normalize the scale factor such that it is unity at the beginning of the scaling regime, 
\beq
\label{equ:ai}
a(\tau_i) \equiv 1\ .
\eeq 
This can always be achieved by a conformal rescaling of the spatial three-metric. 
It will also be helpful to define a dimensionless time variable
\beq
\label{equ:defz}
z \equiv H_i \tau\, ,
\eeq
where $H_i \equiv H(\tau_i)$. 
By a change of variables, $b \equiv 1/a$, equation (\ref{equ:a}) becomes
\beq
\label{condition}
\fbox{$\displaystyle b''= \frac{\alpha}{z^2} (b')^2 b $}\ ,
\eeq
where $'$ now denotes differentiation with respect to $z$ and we have defined the constant $\alpha \equiv c_s \epsilon_i z_i^2$.
Equation (\ref{condition}) is a special case of the {\it generalized Emden-Fowler equation}~\cite{RussianBook}, whose exact solution we present below.
The `slow-roll' parameters derived from $b(z)$ are
\begin{equation}
\hspace{0.7cm} \epsilon = \frac{b'' b}{(b')^2} = \frac{\alpha}{z^2} b^2 \qquad {\rm and} \qquad
\eta = - \frac{(\ln \epsilon)'}{(\ln b)'} = \frac{2}{z} \frac{b}{b'} - 2\ . \label{equ:eta}
\end{equation}
By definition, {cf.}~(\ref{equ:ai}) and (\ref{equ:defz}), the function $b(z)$ satisfes the following boundary conditions:
\beq
\label{in}
b(z_i) = 1 \quad {\rm and} \quad b'(z_i) = -1\ .
\eeq
The solutions to (\ref{condition}) are classified by the two constants $z_i$ and $\epsilon_i$ (or $\alpha$).\,\footnote{From now on we will restrict ourselves to the limit $c_s\approx\,$1, as at this point it is clear that the solution for $c_s\approx const.$ can be trivially recovered from the solution for $c_s = 1$ by substituting $\tau\rightarrow c_s \tau$ and $\alpha\rightarrow c_s\alpha$.}
While $\epsilon_i = \epsilon(z_i)$ characterizes the equation of state at the start of the scaling regime, the variable $z_i =H_i \tau_i$ is the initial ratio between the freeze-out scale, $|\tau_i|$, and the (comoving) Hubble scale, $(a_i H_i)^{-1}$,
\beq
z_i \equiv \frac{\tau_i}{(a_i H_i)^{-1}}=  - 2 \Bigl( \frac{a'}{a}\Bigr)_i \Bigl( \frac{q'}{q}\Bigr)_i^{-1}\ .
\eeq
 In quasi-de Sitter spacetimes the two scales coincide, $z_i = -1-\epsilon_i \approx - 1$, while more generally (i.e.~for time-dependent $\epsilon$) the scales can be quite different.
 We consider the branch of solutions with $\tau_i < 0$ (\footnote{Note that equation (\ref{equ:a}) has a $\mathbb{Z}_2$ symmetry, $z \to - z$, relating the solutions for $\tau_i < 0$ and $\tau_i > 0$.}).
Recalling that $|\tau|$ decreases by at least a factor of $e^{-10}$ during the scaling regime implies $|z_{\rm end}| < e^{-10} |z_i|$.
This exponential change of conformal time during the scaling phase will feature prominently in the strong coupling problem.

\subsection{Approximate Solutions}
\label{sec:approx}

To develop some intuition for the space of solutions to (\ref{condition}), we will first derive approximate solutions as an expansion in small $\alpha$. 
This reproduces three known limiting cases: slow-roll inflation~\cite{TASI}, slow contraction with a  rapidly changing equation of state~\cite{PaulJustin}, and slow expansion with rapidly changing equation of state~\cite{Justin2}.
Then we will show that  (\ref{condition}) in fact permits an {\it exact} solution (\S\ref{sec:exact}).
This will allow us to study FRW backgrounds in full generality (\S\ref{sec:results}).

\vskip 4pt
An iterative solution to (\ref{condition}) as an expansion in powers of $\alpha$ (\footnote{Strictly speaking our solutions will be expansions in $\alpha/ z$ and will break down for small $|z| < \alpha \ll 1$. At that point we will switch to the exact result of \S\ref{sec:exact}.}) may be written as follows:
\beq
b(z) = \sum_{n=0}^\infty  b_{(n)}(z)\, ,
\eeq
where the functions $b_{(n)}(z) = {\cal O}(\alpha^n)$ satisfy
\begin{align}
b''_{(0)} &= 0\ , \\
b''_{(1)} &= \frac{\alpha}{z^2} (b'_{(0)})^2 b_{(0)}\ , \label{b1p}\\
b''_{(2)} &= \frac{\alpha}{z^2} \left[(b'_{(0)})^2 b_{(1)} + 2\, b'_{(0)} b'_{(1)} b_{(0)}\right]\ ,\\
b''_{(3)} &= \frac{\alpha}{z^2}\left[ (b'_{(0)})^2 b_{(2)} + 2\, b'_{(0)} b'_{(2)} b_{(0)} + (b'_{(1)})^2 b_{(0)}  + 2\, b'_{(0)} b'_{(1)} b_{(1)}  \right] \ ,\\
b''_{(4)} &= \ \  {\cdots} \ .
\end{align}
The expansion converges if $b_{(n+1)} \ll b_{(n)}$ and $b'_{(n+1)} \ll b'_{(n)}$.
The initial conditions (\ref{in}) imply
\beq
\label{in2}
b_{(0)}(z_i) = - b_{(0)}'(z_i) = 1 \qquad  {\rm and} \qquad b_{(n \ge 1)}(z_i) = b'_{(n \ge 1)}(z_i) = 0\ .
\eeq
To zeroth-order in $\alpha$, the solution to (\ref{condition}) therefore is
\beq
\label{b0}
b_{(0)} = 1+z_i - z\ .
\eeq
We find the first-order correction $b_{(1)}$ by substituting (\ref{b0}) into (\ref{b1p}), integrating (\ref{b1p}) and imposing (\ref{in2}):
\beq
\label{b1}
b_{(1)} = -
\, \alpha \left[(1+z_i + z) \ln \frac{z}{z_i} + (1+2 z_i) \Bigl(1 - \frac{z}{z_i} \Bigr) \right]\ .
\eeq
Hence, the solution to first-order in $\alpha = \epsilon_i z_i^2$ is
\beq
b \ \approx \ 1 + (z_i - z) \ - \ \epsilon_i z_i^2 \left[(1+z_i + z) \ln \frac{z}{z_i} + (1+2 z_i) \Bigl(1 - \frac{z}{z_i} \Bigr) \right] \ . \label{solution}
\eeq
Continuing this iterative procedure we could determine the solution to arbitrary order in $\alpha$.
However, all the interesting physics is already included at this order of the approximation.
In particular, we will now show that (\ref{solution}) has two limits of particular interest:
slow-roll inflation and slow contraction or expansion with rapidly changing equation of state.

\vskip 6pt
\noindent
{\bf Slow-roll inflation.} \hskip 4pt
In quasi-de Sitter space the ratio of the freeze-out scale to the Hubble scale is 
\beq
\label{equ:zi}
z_i = - 1 - \epsilon_i + {\cal O}(\epsilon_i^2)\ .
\eeq
In this case, the result (\ref{solution}) reduces to
\beq
b \ =\ - z \bigl[ 1 - \epsilon_i + \epsilon_i \ln |z|\, \bigr] \label{inf} \ .
\eeq
Using
\beq
\frac{H}{H_i} = - b' = 1 + \epsilon_i \ln \frac{z}{z_i}\ ,
\eeq
this implies the following solution for the scale factor
\beq
a = b^{-1} = - \frac{1}{H \tau} (1+ \epsilon_i) + {\cal O}(\epsilon_i^2)\ . \label{aaa}
\eeq
This is the conformal time for a quasi-de Sitter space, a familiar result from the standard treatment of inflation~\cite{TASI}. 

We should be careful about the range of validity of the solution (\ref{inf}): 
our convergence criterium,
$b_{(1)} \ll b_{(0)}$ and  $b_{(1)}' \ll b_{(0)}'$, 
will break down for small $z$.  In that limit the stronger condition is $b_{(1)} \ll b_{(0)}$.
Specifically, for $|z| \ll |z_i|$, (\ref{b0}) and (\ref{b1}) become
\beq
 b_{(0)} \approx -\epsilon_i - z  \qquad {\rm and} \qquad  b_{(1)} \approx \epsilon_i \ ,
\eeq
and we hence trust our approximate solution in the regime
$ \epsilon_i < |z| < |z_i|$. In this regime the solution (\ref{inf}) should be a good approximation to the exact answer.

Substituting (\ref{inf}) into (\ref{equ:eta}), we infer the slow-roll parameters
\begin{align}
\label{equ:SReps}
\epsilon &= \epsilon_i (1+ 2 \epsilon_i \ln |z|) + {\cal O}(\epsilon_i^3)\ ,\\
\label{equ:SReta}
\eta &=- 2 \epsilon_i + {\cal O}(\epsilon_i^2)\, .
\end{align}
This result is consistent with the fact that at first-order in a slow-roll expansion the deviation from scale-invariance is given by $n_s - 1 = - 2 \epsilon - \eta$. At first-order in $\epsilon_i$, we indeed find a cancellation between $\epsilon$ and $\eta$: $n_s - 1 = {\cal O}(\epsilon_i^2)$. At second order in $\epsilon_i$ there is a small time-dependence of $\epsilon$ and $\eta$. By construction, this time-dependence is exactly such that scale-invariance is preserved.

Finally, we recall the weak-coupling criterium (\ref{equ:X})
\beq
X = \{ \epsilon, \eta\}\, \zeta_\star \ < \ 1\ .
\eeq
From the time-independence of the couplings (\ref{equ:SReps}) and (\ref{equ:SReta}) we see that the slow-roll solution is indeed weakly coupled for sufficiently long times 
\beq
X(z) \ \ll \ 1 \qquad {\rm for} \quad z_i < z < z_{\rm end}\ .
\eeq

\vskip 6pt
\noindent
{\bf Slow contraction.} \hskip 4pt
For positive $z$ ({i.e.}~negative $H_i$) the result~(\ref{solution}) describes contracting solutions.
Furthermore, in the limit $0< z_i \ll 1$, the solution (\ref{solution}) becomes
\beq
b\ \approx\ 1 + z_i - z - \epsilon_i z_i^2 \left[ \ln \frac{z}{z_i} + \left( 1 - \frac{z}{z_i}\right) \right] \ . \label{bcon}
\eeq
As before, this is a good approximation to the exact solution as long as $b_{(1)} \ll  b_{(0)}$ and $b_{(1)}' \ll  b_{(0)}'$.
Now the stronger condition is $b_{(1)}' < b_{(0)}'$.  Since
\beq
 b_{(0)}'  \approx -1  \qquad {\rm and} \qquad b_{(1)}' \approx - \epsilon_i z_i^2 \frac{1}{z} \ , \\
\eeq
we therefore trust the approximate solution (\ref{bcon}) in the regime
$ \epsilon_i z_i^2 < |z| < |z_i|$.
The first `slow-roll' parameter derived from (\ref{bcon}) grows exponentially with time
\beq
\epsilon  \approx \epsilon_i\, \frac{z_i^2}{z^2}  \, .
\eeq
Essentially, since the scale factor is nearly constant in the solution (\ref{bcon}), all the time-dependence of $q^2 =a^2 \epsilon = \epsilon/b^2 \propto 1/z^2$ is provided by $\epsilon$ rather than by $a$.
This is the {\it contracting adiabatic ekpyrotic phase} first discussed by Khoury and Steinhardt~\cite{PaulJustin}.

Since the time coordinate $z$ changes exponentially -- by at least a factor of $e^{-10}$ during the scaling regime -- one has to worry about the exponential growth of $\epsilon$ as a source of large interactions. Strong coupling can only be avoided if $\epsilon_i$ is exponentially small~\cite{PaulJustin}, in which case there is not enough time for it to blow up before the end of the scale-invariant phase.
However, in this case the second slow-roll parameter typically still leads to trouble:
\beq
\eta  = - \frac{(\ln \epsilon)'}{(\ln b)'} = - \frac{1}{z}  \qquad \Rightarrow \qquad |\eta| \ \ \gtrsim \ \ 1\ .
\eeq
Independent of the size of $\epsilon_i$, the coupling $\eta$ is large and exponentially growing.
In particular, after less than 10 $e$-folds the theory becomes strongly coupled, $X \sim \eta_\star \zeta_\star \sim 1$.

To show that this is a generic problem and not in any significant way tied to our expansion scheme, we consider the time-derivative of the $\epsilon$ parameter in (\ref{equ:eta}):
\beq
\frac{d}{dz} (\, \ln \sqrt{\epsilon}\, ) = - \frac{1}{z} + \frac{b'}{b}\ .
\eeq
Using $b' = - H/H_i$, this can be written as
\beq
\label{equ:epsVar}
\frac{d}{d\tau} (\, \ln \sqrt{\epsilon}\, ) = - \frac{1}{\tau} - \frac{H}{b}\ .
\eeq
During slow-roll inflation with $H = - b/\tau > 0$, a cancellation between the two terms on the r.h.s.~of (\ref{equ:epsVar}) allows for a nearly time-independent $\epsilon$ parameter. However, in a contracting spacetime with $H < 0$ such a cancellation is impossible. Instead, we obtain the inequality~\cite{Justin2}
\beq
\frac{d}{d\tau} (\, \ln \sqrt{\epsilon}\, ) >  \frac{1}{|\tau|}\ . 
\eeq
Hence, during a contracting phase $\epsilon$ is forced to change exponentially, typically leading to a large $\eta$-coupling in the cubic action.

\vskip 6pt
\noindent
{\bf Slow expansion.} \hskip 4pt
Since equation (\ref{condition}) has a symmetry under $z \to - z$, we expect a similar solution, still with $a>0$, corresponding to slow expansion with rapidly changing $\epsilon$. Indeed, for negative $z$ (i.e.~positive $H_i$) the result~(\ref{solution}) describes expanding solutions.
Furthermore, in the limit $|z_i| \ll 1$, the solution reduces to (\ref{bcon}).
This is the {\it expanding adiabatic ekpyrotic phase} discovered by~\cite{Justin2}.
Our above statements concerning the problem of large $\eta$ still apply in this case.

\subsection{Exact Solutions}
\label{sec:exact}

We now show that the non-linear differential equation~(\ref{condition}) in fact has an {\it exact} solution.
Consider first the {\it generalized Emden-Fowler equation}
\beq
\label{equ:EF}
b'' = \alpha\, z^n b^m (b')^l\, .
\eeq
For $\{ n, m, l\} = \{ -2, 1, 2\}$ this is precisely of the form of (\ref{condition}).
Remarkably, this special case has an exact solution~\cite{RussianBook}\,\footnote{Here, we correct a crucial typographical error in the result of \cite{RussianBook}.}\,:
\bea
z(t) &=& \beta_1\, c_1 \left[ G(t) + c_2\right]  \label{zt}\ ,\\
b(t) &=& \beta_2 \, c_1 \left\{ 2 t \left[G(t) +c_2 \right] + \frac{dG}{dt}\right\}\ , \label{b}
\eea
where $\alpha \equiv \frac{1}{2}\beta_1^2 \beta_2^{-2}$ and
\beq
G(t) \equiv \int e^{-t^2} d t = \frac{\sqrt{\pi}}{2} {\rm Erf}(t)\ .
\eeq

\vskip 4pt
\noindent
{\small \underline{Digression}: The following manipulations massage this result into a more user-friendly form:\footnote{Readers without an affinity for these basic algebraic manipulations may jump directly to \S\ref{sec:results} for a graphical representation of the solution.}

\vskip 4pt 
\noindent
First, we may combine (\ref{zt})  and (\ref{b}) into
\beq
b = \beta \left[ 2 t z + \frac{dz}{dt} \right] \ , \qquad {\rm where} \qquad \beta \equiv \frac{\beta_2}{\beta_1} =\pm  \frac{1}{\sqrt{2\alpha}}\ .
\eeq
Furthermore, the first derivative of $b(z)$ is
\beq
\label{equ:bprime}
b' \equiv \frac{db}{dz} =  \frac{db}{dt} \left[ \frac{dz}{dt}\right]^{-1} = 2 \beta z \left[ \frac{dz}{dt}\right]^{-1}\ ,
\eeq
where we used $\frac{d^2 z}{dt^2} = -2 t \frac{d z }{dt}$.
Given $z_i$ and $\epsilon_i$ we can therefore map the initial conditions $b(z_i) = - b'(z_i) = 1$ into the parameters $c_1$ and $c_2$: 
We note from (\ref{equ:bprime}) that $b'(z_i) = -1$ implies
\beq
\left. \frac{dz}{dt} \right|_i = \beta_1 c_1 e^{-t_i^2} = - 2 \beta z_i\ .
\eeq
The constraint $b(z_i) = 1$ then gives
\beq
\label{ti}
t_i = \beta \left[ 1 + \frac{1}{2\beta^2 z_i}\right]\ . 
\eeq
Equation~(\ref{zt})  becomes
\beq
\label{equ:z}
\frac{z}{z_i} = 1 - \gamma \left[ G(t) - G(t_i)\right] \ ,
\eeq
where 
\beq
\label{equ:gamma}
\gamma \equiv - \frac{\beta_1 c_1}{z_i} = 2 \beta \, e^{t_i^2} \ .
\eeq
To implement the solution numerically it is useful to define $x \equiv t- t_i $ and hence obtain
\beq
\gamma [G(t)-G(t_i)] = 2  \beta \, e^{t_i^2} \int_{t_i}^t e^{-t'^2} dt'  = 2 \beta \int_0^x e^{-2 t_i x' - x'^2} dx'\ .
\eeq
Finally, we find
\beq
\fbox{$\displaystyle z(x) = z_i \left[ 1 - 2 \beta  \int_0^x e^{-2 t_i x' - x'^2} dx'  \right] $} \ ,  \label{exactz}
\eeq
\beq
\fbox{$\displaystyle  b(x) = 2 \beta z_i \left[ (t_i + x) \frac{z(x)}{z_i} - \beta \, e^{- 2 t_i x - x^2}\right] $}\ . \label{exactb}
\eeq
These solutions have two input parameters: $z_i$ and $\beta$ (or $\epsilon_i$).
Requiring $|z|$ to decrease with increasing $|x|$ imposes $x>0$ for $\beta > 0$ and $x<0$ for $\beta < 0$.
In fact, recalling that $t_i$ is an odd function of $\beta$, we discover that the solutions have the following symmetry
\beq
z_{\{z_i, \beta\}}(x) = z_{\{z_i, -\beta\}}(-x)\qquad  {\rm and} \qquad b_{\{z_i, \beta\}}(x) = b_{\{z_i, -\beta\}}(-x)\ .
\eeq
Without loss of generality we can therefore focus on the $\beta >0$ branch of the solutions.

From (\ref{exactz}) and (\ref{exactb}) it is straightforward to compute the slow-roll parameters using (\ref{equ:eta}).
With $b_{,x} = 2 \beta z$ we get
\beq
\fbox{$\displaystyle \eta(x) = 2\, \frac{z_i}{z} \frac{b}{(-z)} e^{-2t_i x - x^2} - 2 $}\ .
\eeq }

\subsection{Illustrations of the Strong Coupling Problem}
\label{sec:results}

By construction, fluctuations living in the backgrounds (\ref{exactb}) have scale-invariant two-point correlations.
However, as we have argued above, this does not imply that higher-order correlations are truly scale-invariant. In this section
we use the solution (\ref{exactb}) to analyze the strong coupling problem of \S\ref{sec:strong} in a model-independent way.
We note that our solutions are determined by two input parameters: $z_i$---the ratio of the freeze-out scale to the comoving Hubble scale at the beginning of the scaling phase---and 
$\epsilon_i$---the equation of state at that time.
The parameter $z_i$ may also be viewed as a measure of the deviation of the solutions from the inflationary quasi-de Sitter backgrounds. For inflation we have $z_i = - 1 - \epsilon_i + \epsilon_i^2$, while backgrounds with $z_i$ far from $-1$ are very different from slow-roll inflation.

In this section, we will scan over the possible values of the input parameters $z_i$ and $\epsilon_i$. 
As  a diagnostic for the strong coupling problem we will then compute the $\eta$ parameter $N$ $e$-folds after the beginning of the scaling regime:
\beq
b_{\{z_i, \epsilon_i\}}(z) \quad \Rightarrow \quad \eta_{N} \ ,
\eeq
where $z = e^{- N} z_i$. For $\log \eta_{N} \gtrsim 4$ we lose perturbative control. Even before that, for $\log \eta_{N} \gtrsim 2$, we expect tension with constraints on non-Gaussianity.


\vskip 6pt
\noindent
{\bf Slow-roll inflation.} \hskip 4pt This is the most well-known case. We have perturbative control for an arbitrary amount of time. We have confirmed numerically that this regime is accurately described by the approximate solution (\ref{inf}) (within its expected range of validity) and the exact solution (\ref{exactz}) and (\ref{exactb}). For the exact solution to produce more than 60 $e$-folds may require $z_i$ to be defined to second order in $\epsilon_i$.

\begin{figure}[h!]
    \centering
        \includegraphics[width=.6\textwidth]{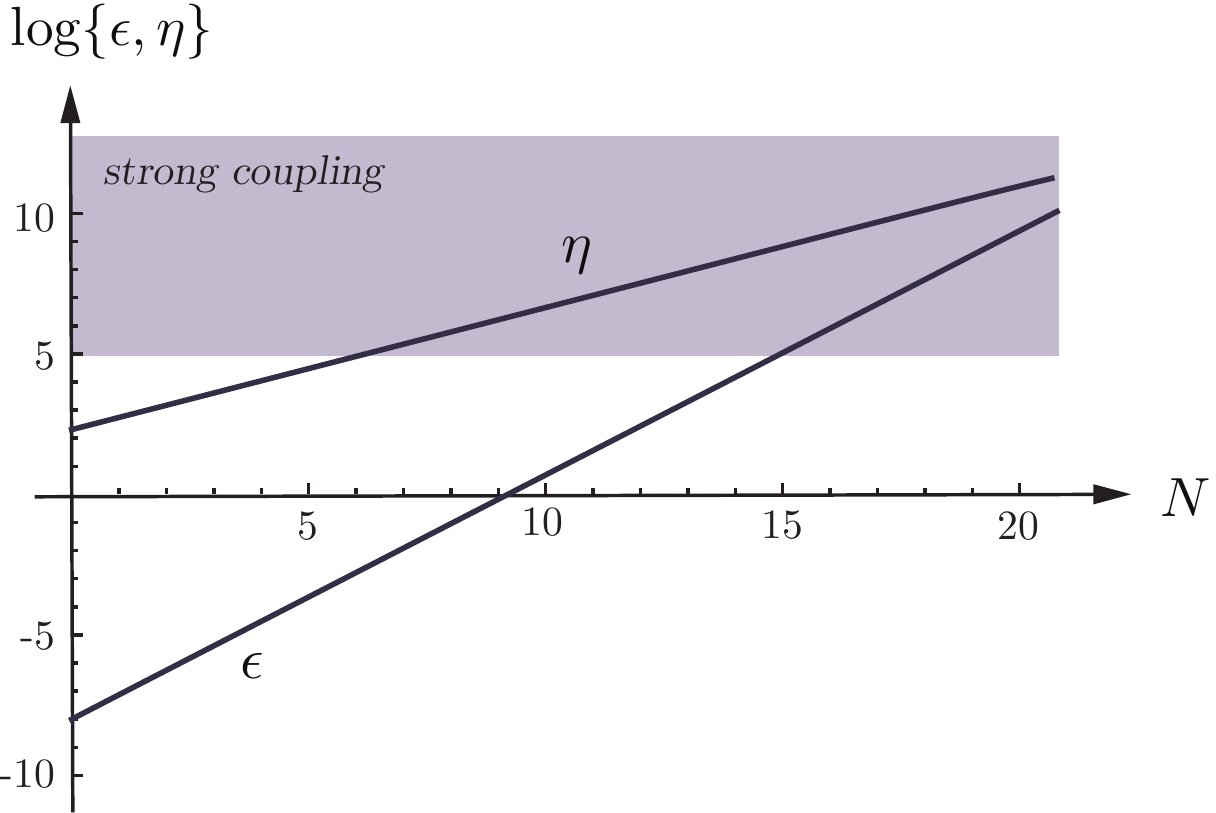}
    \caption{\sl Slow contraction: plot of $\log \epsilon$ and $\log \eta$ vs.~$N$ for $\epsilon_i = 10^{-8}$ and $z_i = 0.01$.}
    \label{fig:EpsNCon}
\end{figure}

\vskip 6pt
\noindent
{\bf Slow contraction.} \hskip 4pt
In contrast, in Figure~\ref{fig:EpsNCon} we show a canonical example of a slowly contracting spacetime with $\epsilon_i = 10^{-8}$ and $z_i =0.01$. 
As expected, $\epsilon$ and $\eta$ grow exponentially and within a small number of $e$-folds the theory becomes strongly coupled. At this point perturbative control is lost and, in the absence of a UV-completion or a proof of strongly coupled unitarization, the theory becomes unpredictive.

\begin{figure}[h!]
    \centering
        \includegraphics[width=.9\textwidth]{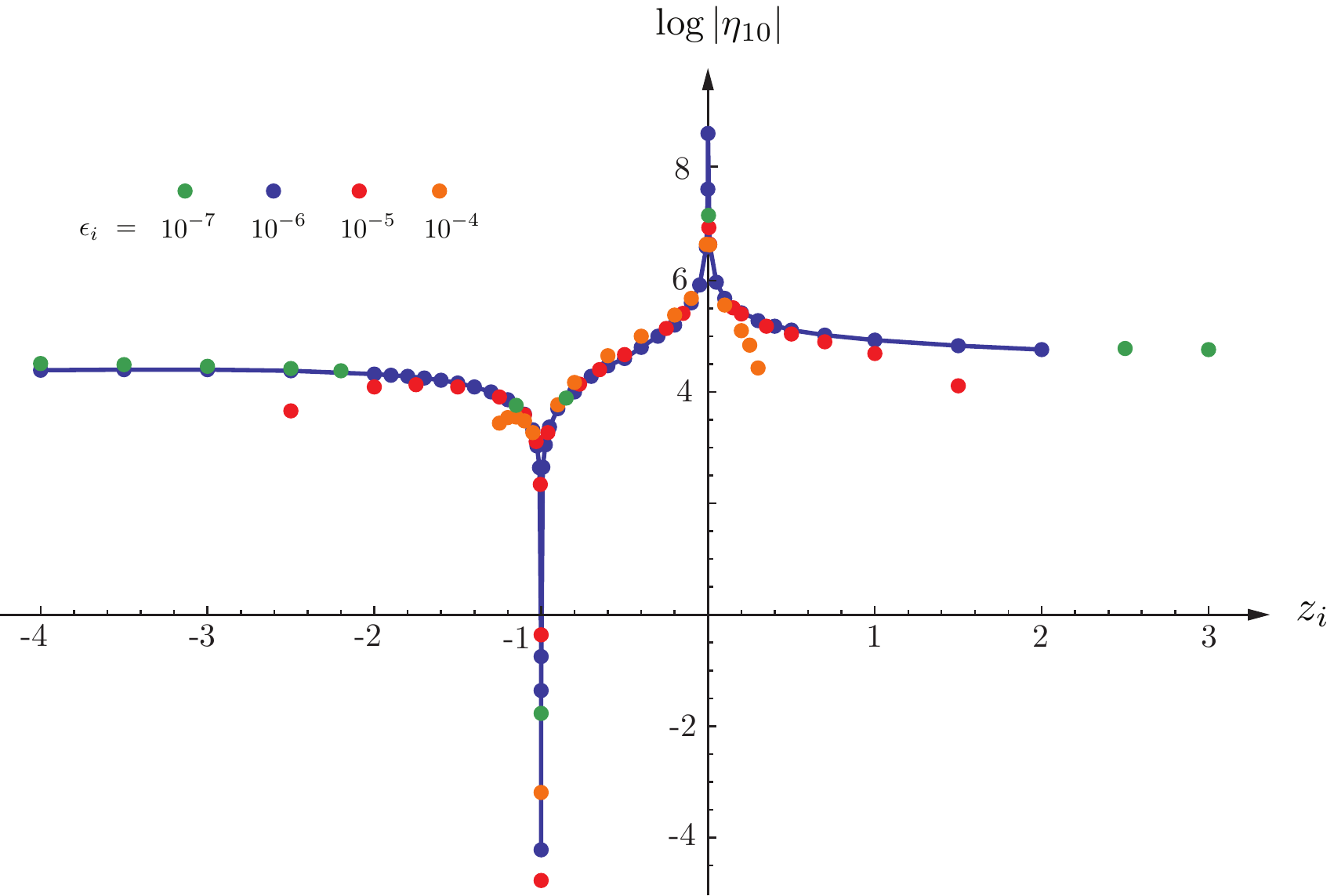}
    \caption{\sl The value of the $\eta$ parameter 10 $e$-folds after the beginning of the scaling phase illustrates the strong coupling problem arising in backgrounds that deviate too much from quasi-de Sitter ($z_i = - 1 - \epsilon_i$).}
    \label{fig:ziScan}
\end{figure}

\vskip 6pt
\noindent
{\bf de Sitter vs.~non-de Sitter.} \hskip 4pt
In Figure~\ref{fig:ziScan} we present a more general scan of the parameter space of solutions. For fixed $\epsilon_i = 10^{-4}, 10^{-5}, 10^{-6}, 10^{-7}$ we scan the parameter $z_i$.
We plot the maximal value of $\eta$ in the interval $[z_i, z_{\rm end}= e^{-10} z_i]$. 
We find numerically that away from $z_i \approx -1$ or $|z_i| \ll 1$ it is hard to produce even just 10 $e$-folds of scale-invariant modes unless $\epsilon_i$ is very small (see also \cite{Justin2}). This is the reason that the points corresponding to $\epsilon_i = 10^{-4}$ and $10^{-5}$ don't extend to larger values of $|z_i|$. Only for the range of $z_i$ shown does the background produce at least 10 $e$-folds of modes.
For $z_i \approx - 1$ or $|z_i| \ll 1$ the background allows more than 10 $e$-folds even without exponentially small $\epsilon_i$.

The plot includes all three phases mentioned above:
{\it i})~slow-roll inflation ($z_i \approx - 1 - \epsilon_i + \epsilon_i^2$), {\it ii})~slow-contraction ($0< z_i \ll 1$) and {\it iii})~slow-expansion ($-1 \ll z_i < 0$).
We see that all backgrounds except quasi-de Sitter suffer from a strong coupling problem, making it highly unlikely that more than 10 $e$-folds of scale-invariant modes can be realized in a controlled way.
If one is satisfied with just producing the modes observed in the CMB, one may consider creating $\sim 5$ $e$-folds of scale-invariant modes in a non-de Sitter evolution.
To decide whether these modes would be consistent with observational constraints on non-Gaussianity would require a more careful computation of the three-point function including the effects of other interactions that could partially cancel the large contribution from the interaction proportional to $\eta$. Why those few $e$-folds would happen to be the ones that we have observational access to would require an explanation.

\subsection{Generalizations}
\label{sec:general}

So far our analysis was limited to actions with trivial propagation speed for the cosmological curvature perturbations, $c_s= const$.
In this subsection, we offer a few comments on extensions of our results to theories with non-trivial speed of sound and to cases in which high-derivative terms play an important role in the dynamics.

\vskip 6pt
\noindent
{\bf Varying speed of sound.} \hskip 4pt  For attractor solutions, the requirement of weak coupling at horizon crossing implies a lower limit on the sound speed
\beq\label{eq:csinteraction}
\frac{{\cal L}_3}{{\cal L}_2} = \left. \frac{\zeta}{c_s^2} \right|_\star\ \lesssim \ 1 \qquad \Rightarrow \qquad c_{s, \star}\ \gtrsim\ \zeta_\star^{1/2} \sim 10^{-3}\ .
\eeq
The observational upper limit on the level of non-Gaussianities in the CMB strengthens this bound by one or two orders of magnitudes~\cite{SenatoreSmithWMAP5}.\footnote{As we stressed before, having a small, constant speed of sound is not much of an issue once the constraint from non-Gaussianities is satisfied.}
Moreover, in order to avoid superluminality the sound speed is bounded from above, $c_s \le 1$. Hence, the sound speed can vary at most by one or two orders of magnitude over the entire range of the scaling regime from $y_i$ to $y_{\rm end} = e^{- \Delta N} y_i$.
The allowed time-variation of the speed of sound is therefore too constrained to have a significant effect on $q^2 \propto 1/c_s$.
In particular, we do not believe that allowing the speed of sound to be time-dependent will give qualitatively new, non-de Sitter solutions.

If we are looking for a scale-invariant solution that can last at least 60 $e$-foldings, then the interactions proportional to $\epsilon$ and $\eta$ will force us to have $\epsilon$ approximately constant. In this regime, the requirement for scale invariance becomes:
\beq
 \frac{a^2 \epsilon}{c_s} = \frac{\epsilon_i}{c_{s, i}} \Bigl(\frac{y_i}{y}\Bigr)^2 \qquad\Rightarrow\qquad c_s\propto (ay)^2\ .
\eeq
This allows us to derive $a(y)$ as a function of $y$:
\beq
\epsilon=-\frac{\dot H}{H^2} = - \frac{a'' a}{(a')^2}- \frac{2}{y}\frac{a}{a'}  \quad\Rightarrow\quad a=a_i\Bigl(\frac{y_i}{y}\Bigr)^{\frac{1}{1+\epsilon}}\ .
\eeq
Plugging this into (\ref{eq:csinteraction}), we get 
\beq
\frac{{\cal L}_3}{{\cal L}_2} = \left. \frac{\zeta}{c_s^2} \right|_\star\sim \Big(\frac{y_i}{y}\Big)^{\frac{4\epsilon}{1+\epsilon}}\, \zeta \ .
\eeq
To avoid that the interactions grow catastrophically as $y\rightarrow 0$, we require $0<\epsilon\ll 1$ or $\epsilon<0$. The second option makes the fluctuations ghost-like, so we are left with inflation as the only weakly-coupled solution.


\vskip 6pt
\noindent
{\bf Higher-derivative terms.} \hskip 4pt
In Appendix~\ref{sec:A2} we derive the most general quadratic action for curvature fluctuations $\zeta$ in the effective field theory approach of \cite{Creminelli:2006xe,Cheung}.
In unitary gauge there are no matter fluctuations, but only metric fluctuations, so we define the effective action by writing down all operators that are functions of the metric fluctuations and invariant under time-dependent spatial diffeomorphisms.
The two most important elements appearing in this construction are the metric perturbation $\delta g^{00}$ and the extrinsic curvature perturbation $\delta K_{ij}$.
In Appendix~\ref{sec:A2} we use these geometrical quantities to define the most general action with unbroken spatial diffeomeophisms.
As usual in effective field theories, this can be done in a low-energy expansion of the fields and their derivatives: $\delta g^{00}$ is scalar with zero derivatives acting on it, while $\delta K_{ij}$ is a one-derivative object. 
In many situations the terms involving $\delta g^{00}$ therefore dominate the dynamics.
This leads to a quadratic action for $\zeta$ with non-trivial sound speed and is hence captured by the analysis in this section.
However, in some limits, identified by a particular symmetry structure of the Lagrangian for the fluctuations, the terms involving the extrinsic curvature $\delta K_{ij}$ can in fact become the dominant effects.
For instance, in the limit in which $M_{\rm pl}^2 \dot H$ is smaller than a combination of the coefficients of the extrinsic curvature terms and the Hubble parameter $H$ (see Appendix~\ref{sec:new}), the extrinsic curvature terms can lead to a de Sitter background with significant higher-derivative terms in the action for $\zeta$. These are the {\it ghost inflation}~\cite{GhostInflation} regime and the solutions identified as `near-de-Sitter limit' in~\cite{Cheung} and \cite{SenatoreSmithWMAP5}.

Can there be non-de Sitter backgrounds in which higher-derivative terms allow weakly-coupled scale-invariant fluctuations?
In Appendix~\ref{sec:new} we address this question.  Although we don't give a comprehensive analysis of the range of possibilities implied by the rather complicated actions for $\zeta$, we provide an extensive discussion of the most interesting limits of the parameter space. We reproduced the known near-de Sitter solutions, but found no non-de Sitter backgrounds with weakly-coupled scale-invariant fluctuations.

\section{Summary}
\label{sec:conclusions}

Cosmic microwave background (CMB) and large-scale structure (LSS) observations constrain the primordial seed fluctuations to be {\it i}) nearly scale-invariant and {\it ii}) approximately Gaussian (or weakly coupled).
In this paper we asked what these two observational facts together teach us about the cosmological background at the time when these fluctuations exited the horizon.
We considered the most general effective theory of adiabatic fluctuations around completely arbitrary FRW backgrounds allowing both for expansion and contraction. We assumed that the observed cosmological perturbations arose from quantum mechanical fluctuations of a single degree of freedom. 
For theories with constant sound speed and requiring the background to be a dynamical attractor, 
we reduced the requirement of scale-invariant two-point correlations to a non-linear differential equation for the scale factor $a(\tau)$,
\beq
q^2 = a^2 \epsilon = a^2 \left[ 2 - \frac{a'' a}{(a')^2}\right] \propto \frac{1}{\tau^2}\ .
\eeq
This equation has an exact solution---a special case of solutions to the generalized Emden-Fowler equation~\cite{RussianBook}. Using this solution we reproduced two known limits: slow-roll inflation~\cite{TASI} ($a \propto \tau^{-1}$) and ekpyrotic contraction (or expansion) with rapidly changing equation of state~\cite{PaulJustin,Justin2} ($\epsilon \propto \tau^{-2}$) .
  \begin{figure}[h!]
    \centering
        \includegraphics[width=.8\textwidth]{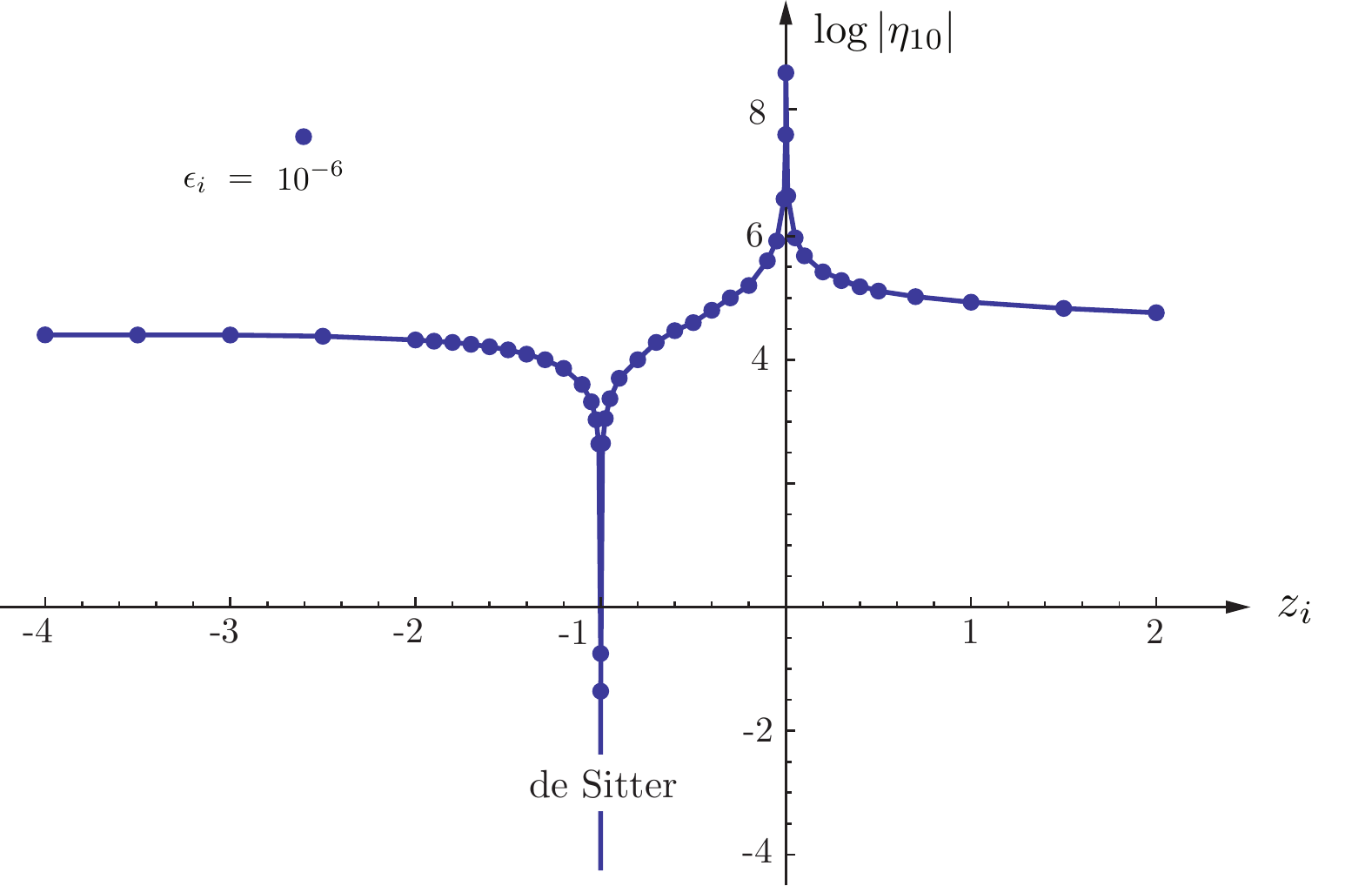}
    \caption{\sl The value of the $\eta$ parameter 10 $e$-folds after the beginning of the scaling phase illustrates the strong coupling problem arising in backgrounds that deviate too much from quasi-de Sitter ($z_i = - 1 - \epsilon_i$).}
    \label{fig:ziScan2}
\end{figure}

The non-linear realization of time diffeomorphisms in the effective theory of the fluctuations forces specific relationships between the coefficients of the quadratic Lagrangian and the interaction Lagrangian. This drastically limits the number of viable models that can produce a scale-invariant two-point function while staying weakly coupled for a sufficiently long time.
In fact, we showed that only inflation leads to nearly time-independent couplings of all higher-order interactions such as $\epsilon$, $\eta$ and $c_s^{-2}$. This implies that only inflationary spacetimes allow the fluctuations to stay weakly coupled over the range of scale relevant to cosmological observations.
For all non-de Sitter backgrounds we identified a strong coupling problem that limits the predictivity of these solutions.
Within 10 $e$-folds after the beginning of the scaling phase the perturbative expansion of the theory breaks down. This makes it challenging to produce the modes observed in the CMB and the LSS in a consistent non-de Sitter background.

Dropping the requirement that the background should be an attractor, we found non-de Sitter solutions with rapidly time-varying speed of sound (following previous work by Khoury and Piazza~\cite{Justin}).
In Appendix~\ref{sec:nonA} we identified the regime of parameter space in which those theories evade both quantum mechanical and classical strong coupling problems.
Since these backgrounds are not attractors, the curvature perturbations $\zeta$ evolve after horizon exit and predictions depend on assumptions about the physics both before and after the regime during which the scale-invariant modes exit the horizon.
We described the basic model-building requirements for non-attractor non-de Sitter backgrounds with weakly-coupled scale-invariant fluctuations. 
Whether these physical elements can be realized in a coherent theoretical framework remains an open question.
In contrast, it is quite remarkable how the time-translation invariance of quasi-de Sitter backgrounds without any additional physical ingredients solves the horizon and flatness problems, while at the same time allowing for scale-invariant $n$-point functions.

\vskip 8pt
{\it Note added.} While this paper was being completed, Ref.~\cite{Justin2} appeared which has some overlap with our Section~\ref{sec:attractor}.

\subsubsection*{Acknowledgements}

D.B.~thanks Justin Khoury, Federico Piazza and Paul Steinhardt for helpful discussions and correspondence. 
The research of D.B.~is supported by the National Science Foundation under PHY-0855425 and a William D.~Loughlin Fellowship at the Institute for Advanced Study.
L.S.~is supported in part by the National Science Foundation under PHY-0503584.
M.Z.~is supported by the National Science Foundation under PHY-0855425 and AST-0907969, as well as by the David and Lucile Packard Foundation and the John D.~and Catherine T.~MacArthur Foundation.

\newpage
\appendix

\section{Quadratic Action for $\zeta$}
\label{sec:A2}

In this appendix we derive the second-order Lagrangian for scalar curvature fluctuations $\zeta$ around arbitrary FRW backgrounds, including the mixing of the scalar with gravity~\cite{Creminelli:2006xe, Cheung}.
We start in unitary gauge, in which there are no matter fluctuations, but only metric fluctuations.
The most general effective action is then constructed by writing down all operators that are functions of the metric fluctuations and invariant under time-dependent spatial diffeomorphisms. Time diffeomorphisms are broken by the time-dependence of the background.
For this purpose it is particularly convenient to work in the ADM formalism~\cite{ADM} since it keeps the invariance under spatial diffeomorphisms manifest: only quantities that are covariant under spatial transformations appear in the equations.
The perturbed four-dimensional metric in ADM variables is
\beq
\d s^2 = - N^2 \d t^2 + h_{ij} (\d x^i + N^i \d t) (\d x^j + N^j \d t)\ ,
\eeq
where $N({\bf x}, t)$ and $N_i({\bf x}, t)$ are the `lapse' and `shift' functions, respectively, and $h_{ij}$ is the induced metric on three-dimensional hypersurfaces of constant time $t$.
The geometry of these hypersurface is characterized by the
intrinsic curvature $R^{(3)}_{ij}$, {i.e.}~the Ricci tensor of the induced 3-metric, and the extrinsic curvature $K_{ij}$:
\beq
E_{ij} \equiv N K_{ij} = \frac{1}{2} \left[ \dot h_{ij} - \nabla_i N_j - \nabla_j N_i \right]\ ,
\eeq
where $\nabla$ is the covariant derivative associated with $h_{ij}$.
We will use these geometrical quantities to define the most general action with unbroken spatial diffeomeophisms.
As usual in effective field theories, this can be done in a low-energy expansion of the fields and their derivatives:
We notice that $N$ is a scalar with zero derivatives acting on it, $E_{ij}$ is a one-derivative object and $R^{(3)}_{ij}$ contains two derivatives.

\subsection{Zero-Derivative Action}

At the zero-derivative level in ADM variables the only quadratic operator is $(\delta N)^2$.
Let us therefore study the action
\beq
\label{equ:action1}
S =  \int \d^4 x \sqrt{-g} \left[\frac{M_{\rm pl}^2}{2} R^{(4)} - M_{\rm pl}^2 \left( \frac{1}{N^2} \dot H + 3 H^2 + \dot H \right)  +  M^4(t) (\delta N)^2 \right]\ .
\eeq
where $R^{(4)} = R^{(3)} + N^{-2} (E^{ij} E_{ij} - E^i_{\, i}{}^2)$.
We fix spatial diffeomorphisms by the gauge choice
\beq
h_{ij} = a^2(t) (1 + 2 \zeta) \delta_{ij}\ .
\eeq
The Einstein constraint equations express $N$ and $N_i$ in terms of $\zeta$.
In particular, variation of the action with respect to $N$ gives the Hamiltonian constraint
\beq
\label{equ:Hcon}
\frac{M_{\rm pl}^2}{2} \left(R^{(3)}  - \frac{1}{N^2} (E^{ij} E_{ij} -E^i_{\, i}{}^2) \right) - M_{\rm pl}^2 \left( - \frac{1}{N^2} \dot H + 3 H^2 + \dot H \right) +  2 M^4 \delta N = 0\ ,
\eeq
while the variation with respect to $N^i$ gives the momentum constraint
\beq
M_{\rm pl}^2 \nabla_j \left( \frac{1}{N} (E^j_{\, i} - \delta^j_i E^k_{\, k})\right) = 0\ .
\eeq
To get the quadratic action we only have to solve these constraints at first order in $\zeta$.
Defining $N= 1+N_1$ and $N^i = \partial_i \psi$, we find
\bea
N_1 &=& \frac{\dot \zeta}{H}\ , \\
\nabla^2 \psi &=& \left( \frac{2M^4}{M_{\rm pl}^2 H^2} - \frac{\dot H}{H^2} \right) \dot \zeta - \frac{\nabla^2 \zeta}{H a^2}\ .
\eea
Substituting this back into the action \ref{equ:action1}, we get
\beq
S_2 = M_{\rm pl}^2 \int \d^3 x\, \d t\, a^3(t) \left[ A(t) \dot \zeta^2 + B(t) \frac{(\partial_i \zeta)^2}{a^2} \right]\, ,
\eeq
where
\beq
A(t) \equiv \frac{\epsilon}{c_s^2} \qquad {\rm and} \qquad
B(t) \equiv - \epsilon = \frac{\dot H}{H^2}\ .
\eeq
The dynamics of $\zeta$ is hence determined by two functions of time: the Hubble expansion rate $H(t)$ and the 
 `speed of sound'
\beq
\frac{1}{c_s^2} \equiv 1 - \frac{2 M^4}{ M_{\rm pl}^2 \dot H}\ .
\eeq
Defining a new time variable $y$ via $\d y =(c_s/a) \d t$, the action becomes
\beq
\label{equ:mainA}
\fbox{$\displaystyle S_2 = M_{\rm pl}^2 \int \d^3 x \, \d y \, q^2 \left[ (\zeta')^2 - (\partial_i \zeta)^2 \right] $} \ , \qquad {\rm where} \qquad \fbox{$\displaystyle q^2 \equiv \frac{a^2 \epsilon}{c_s} $}\ .
\eeq
This action is the starting point for much of our discussion in the main text of this paper.

\subsection{Higher-Derivative Terms}

At next order in the derivative expansion we must include terms that involve the extrinsic curvature $E_{ij}$.
We will be interested in fluctuations around the background $\bar E_{ij} = a^2 H \delta_{ij}$, i.e.
\beq
\delta E_{ij} = E_{ij} - \bar E_{ij}\ .
\eeq
There are now three additional quadratic operators: $\delta N \delta E^i_{\, i}$, $\delta E^{ij} \delta E_{ij}$ and $(\delta E^{i}_{\, i})^2$. In order to simplify the calculations, we will treat the operators one at a time.

\subsection*{$(\delta E^{i}_{\, i})^2$ Term}

Let us first add the operator $(\delta E^{i}_{\, i})^2$ to the action (\ref{equ:action1}),
\beq
S =  \int \d^4 x \sqrt{-g} \left[\frac{M_{\rm pl}^2}{2} R^{(4)} - M_{\rm pl}^2 \left( \frac{1}{N^2} \dot H + 3 H^2 + \dot H \right)  +  M^4(t) (\delta N)^2 - \bar M^2(t) (\delta E^i_{\, i})^2 \right]\ .
\eeq
The Hamiltonian constraint is the same as in (\ref{equ:Hcon}), while the momentum constraint becomes
\beq
M_{\rm pl}^2 \nabla_j \left( \frac{1}{N} (E^j_{\, i} - \delta^j_i E^k_{\, k})\right) - 2 \bar M^2 \nabla_i E^k_{\, k} = 0\ .
\eeq
The first order solutions to the constraint equations now are
\bea
N_1 &=& \frac{M_{\rm pl}^4 H \cdot \dot \zeta +  M_{\rm pl}^2 \bar M^2 \cdot \nabla^2 \zeta/a^2}{M_{\rm pl}^4 H^2 + \bar M^2 M^4}\ , \\
\nabla^2 \psi &=& \frac{(-9 M_{\rm pl}^2 \bar M^2 H^2 +  M_{\rm pl}^2 M^4 -  M_{\rm pl}^4 \dot H) \dot \zeta -  M_{\rm pl}^4 H \cdot \nabla^2 \zeta/a^2}{M_{\rm pl}^4 H^2 + \bar M^2 M^4}\ .
\eea
Substituting this back into the action we get~\cite{Creminelli:2006xe}
\beq
\label{equ:ghost}
S_2 = M_{\rm pl}^2 \int \d^3 x\, \d t\, a^3(t) \left[ \bar A(t) \dot \zeta^2 + \bar B(t) \frac{(\partial_i \zeta)^2}{a^2} + \bar C(t) \frac{(\partial^2 \zeta)^2}{a^4} \right]\, ,
\eeq
where
\bea
\bar A(t) &\equiv& \frac{ M_{\rm pl}^2 (M^4 -  9 \bar M^2 H^2 - M_{\rm pl}^2 \dot H)}{ M_{\rm pl}^4 H^2 + \bar M^2 M^4}\ , \\
\bar B(t) &\equiv& \frac{1}{(M_{\rm pl}^4 H^2 + \bar M^2 M^4)^2} \Bigl[ - 3 M_{\rm pl}^6 \bar M^2 H^4 + \bar M^2 (M^4-  M_{\rm pl}^2 \dot H) (M^4 \bar M^2 -  M_{\rm pl}^4 \dot H)\nonumber \\
&& \hspace{1cm} H^2 (M^4 M_{\rm pl}^4 \bar M^2 +  M_{\rm pl}^8 \dot H) -  M_{\rm pl}^6 \bar M^2 H \ddot H \Bigr]\ ,\\
\bar C(t) &\equiv& - \frac{ M_{\rm pl}^2 \bar M^2}{ M_{\rm pl}^4 H^2 + \bar M^2 M^4}\ .
\eea
In Appendix~\ref{sec:new} we will analyze certain interesting limits of this action.

\subsection*{$\delta N \delta E^i_{\, i}$ Term}

Next, we add the operator $\delta N \delta E^i_{\, i}$ to the action (\ref{equ:action1}),
\beq
S =  \int \d^4 x \sqrt{-g} \left[\frac{M_{\rm pl}^2}{2} R^{(4)} - M_{\rm pl}^2 \left( \frac{1}{N^2} \dot H + 3 H^2 + \dot H \right)  +  M^4(t) (\delta N)^2 - \hat M^3(t) \delta N \delta E^i_{\, i} \right]\ .
\eeq
Performing a similar analysis as before we find~\cite{Creminelli:2006xe}
\beq
\label{equ:ghost2}
S_2 = M_{\rm pl}^2 \int \d^3 x\, \d t\, a^3(t) \left[ \hat A(t) \dot \zeta^2 + \hat B(t) \frac{(\partial_i \zeta)^2}{a^2} \right]\, ,
\eeq
where
\bea
\hat A(t) &\equiv& \frac{-12 M_{\rm pl}^2 H \hat M^3 + 4 M^4 M_{\rm pl}^2 - 4 M_{\rm pl}^4 \dot H}{(-\hat M^3 + 2 M_{\rm pl}^2 H)^2}\ , \\
\hat B(t) &\equiv&  \frac{\hat M^6 -2 M_{\rm pl}^2 H \hat M^3 + 4 M_{\rm pl}^4 \dot H}{(-\hat M^3 + 2 M_{\rm pl}^2 H)^2} \ .
\eea
This action is of the same form as (\ref{equ:mainA}) but with a different normalization and a sound speed given by
\beq
\hat c_s^2 \equiv - \frac{\hat B}{\hat A} = \frac{\hat M^6 - 2 M_{\rm pl}^2 H \hat M^3 + 4 M_{\rm pl}^4 \dot H}{12 M_{\rm pl}^2 H \hat M^3 - 4 M^4 M_{\rm pl}^2 + 4 M_{\rm pl}^4 \dot H}\ .
\eeq
Notice that the action is defined by {\it three} functions of time: $a(t)$, $M(t)$ and $\hat M(t)$.

\section{Scale-Invariance from Higher-Derivative Terms?}
\label{sec:new}

Could the actions (\ref{equ:ghost}) and (\ref{equ:ghost2}) allow for scale-invariant fluctuations in non-de Sitter backgrounds without leading to a strong coupling problem? 
The goal of this appendix is to address this question.
Since a comprehensive analysis of the full range of possibilities is beyond the scope of this paper, we will focus on a number of interesting limiting cases. 

\subsection*{$\delta N \delta E^i_{\, i}$ Term}

For notational convenience, we define the following dimensionless ratios
\beq
X\equiv \frac{M^4}{M_{\rm pl}^2 H^2} \qquad {\rm and} \qquad Y \equiv \frac{1}{2} \frac{\hat M^3 H}{M_{\rm pl}^2 H^2}\ ,
\eeq
where, despite appearance, $X$ and $Y$ can both be negative.
In this parameterization we find
\bea
\hat A &=& \bigl(\epsilon + X - 6Y \bigr) \bigl(1 - Y \bigr)^{-2} \ , \label{equ:hatA}\\
\hat c_s^2  &=&  \bigl(\epsilon + Y  - Y^2 \bigr) \bigl(\epsilon + X - 6Y   \bigr)^{-1}  \ . \label{equ:hatcs2}
\eea
We assume $M \sim \hat M$ and $H \ll M$, so that $|X| \gg |Y|$. 
We will furthermore restrict the discussion to the following two limits:
\begin{enumerate}
\item[{\it i})]  $|M^4| \ll M_{\rm pl}^2| \dot H|$ \quad $\Leftrightarrow$ \quad $|X| \ll \epsilon$\ , 
\item[{\it ii})] $|M^4| \gg M_{\rm pl}^2 |\dot H|$  \quad $\Leftrightarrow$ \quad  $|X| \gg \epsilon$\ .
\end{enumerate}
In the first case, $|Y| \ll |X| \ll \epsilon$, (\ref{equ:hatA}) and (\ref{equ:hatcs2}) simplify to
\beq
\label{equ:B4}
\hat A \ \to\ \epsilon\, \bigl(1- Y\bigr)^{-2} \qquad {\rm and} \qquad
\hat c_s^2 \ \to\ 1\ .
\eeq
For $|Y| \ll 1$, this reproduces the canonical action of \S\ref{sec:attractor}: $\hat A \to \epsilon$. In this case, we know that only slow-roll inflation leads to a weakly-coupled solution.
For $|Y| \gtrsim 1$, the solution differs from slow-roll inflation, but it seems hard to avoid the strong coupling problem from interactions proportional to $\epsilon \gg |X| > |Y| \gtrsim 1$. Nevertheless, it is possible to find a scale-invariant attractor solution by choosing $1 - Y \propto \tau$, while keeping $\epsilon$ large and time-independent. This would realize a contracting ekpyrotic universe. However, in this case metric perturbations $\delta N \propto (1 - Y)^{-1} \propto \tau^{-1}$ become large, and interactions mediated by the mixing with gravity become strongly coupled. 

The more interesting limit therefore is the second case, $|X| \gg \epsilon$, for which
we obtain
\bea
\hat A\ \to\  X \bigl( 1- Y \bigr)^{-2} \label{equ:AA} \qquad {\rm and} \qquad
\hat c_s^2 \ \to\ \frac{1}{X}  \bigl(\epsilon + Y - Y^2 \bigr) \ .
\eea
To avoid a ghost instability we require $\hat A > 0$ and hence $X > 0$.
For $|Y| < {\rm min}\{1, \epsilon\}$, the extrinsic curvature corrections proportional to $Y$ are subdominant and the theory reduces to
\bea
\hat c_s^2 \ \to\ \frac{\epsilon}{X} \ll 1 \qquad {\rm and} \qquad  \hat A \ \to\ X = \frac{\epsilon}{\hat c_s^2}   \ .
\eea
This is the case that we studied in the main text and which produced inflation as the only weakly-coupled solution.

The next parameter region to consider is $1 \lesssim |Y| \ll \epsilon \ll |X|$.
The coefficients of the quadratic action become
\beq
\hat c_s^2\ \rightarrow\ \frac{\epsilon}{X} \ll 1  \qquad {\rm and} \qquad  \hat A\ \rightarrow\ \frac{X}{(1-Y)^2}= \frac{\epsilon}{\hat c_s^2} \bigl( 1- Y\bigr)^{-2} \ .
\eeq
In this limit, interactions proportional to $X$ have to be controlled to avoid a strong coupling problem. For instance, a particularly dangerous interaction is
\beq
\frac{{\cal L}_{\zeta (\partial_i \zeta)^2}}{{\cal L}_{\dot \zeta^2}} \sim \frac{\epsilon}{\hat c_s^2} (1- Y)^2 \, \zeta\ .
\eeq
We require $Y$ to be sufficiently different from 1, in order to avoid the strong coupling problem discussed below Eqn.~(\ref{equ:B4}).
Since $\epsilon$ is bounded from below, $\epsilon \gg 1$, and $\hat c_s$ is bounded from above, $\hat c_s \le 1$, we cannot afford much time-variation of both $\epsilon$ and $\hat c_s$ while keeping the fluctuations weakly coupled.
We therefore consider $\epsilon \simeq const$ and $\hat c_s \simeq const$.
The condition for scale invariance then implies
\beq
\label{equ:SInv}
a^2 \hat A \, \hat c_s = \frac{a^2 \epsilon}{\hat c_s} \bigl(1 - Y \bigr)^{-2} \propto {y}^{-2} \propto \tau^{-2} \ .
\eeq
For constant $\epsilon$, we get a power law solution for $a(\tau)$,
\beq
\epsilon = 2 - \frac{a'' a}{(a')^2} \qquad \Rightarrow \qquad a \propto \tau^{1/(\epsilon-1)}\ ,
\eeq
and hence
\beq
1 - Y \propto \tau^{1+ \frac{1}{\epsilon-1}} \approx \tau\ .
\eeq
This quickly leads to a strong coupling problem as $Y$ approaches 1---see the discussion below~(\ref{equ:B4}).

The last case we need to discuss is: $1 > |Y| \gg \epsilon$~(\footnote{We disregard the case $|Y| >1$ because it leads to  $\hat c_s^2 < 0$, signaling that this limit is generically plagued by gradient instabilities. These instabilities may be cured by adding the $(\delta E^i_i)^2$ operators, but only at the expense of having the two higher-derivative terms be equally important near the horizon scale, as described in~\cite{SenatoreSmithWMAP5}. While this is quite a reasonable and technically justified assumption in the case where all the parameters are approximately time-independent as in inflation~\cite{SenatoreSmithWMAP5}, we do not consider this an interesting scenario when all of the parameters are strongly time-dependent.}),
\beq
\hat c_s^2 \ \to\  \frac{Y}{X} \ll1  \qquad {\rm and} \qquad
\hat A \ \to\ X = \frac{Y}{\hat c_s^2} \ . \label{equ:hatcs}
\eeq
As before, interactions proportional to $\hat c_s^{-2}$ restrict us to solutions with $\hat c_s \simeq const$.
Scale invariance then requires
\beq
a^2\hat A \, \hat c_s \approx  \frac{a^2 Y}{\hat c_s}  \propto \tau^{-2} \qquad \Rightarrow\qquad Y \propto X = \frac{M^4}{M_{\rm pl}^2 H^2} \propto \frac{1}{(a\tau)^2} \ , \label{equ:last}
\eeq
For $a \tau = const$, this reproduces the near-de Sitter solutions of~\cite{Cheung,SenatoreSmithWMAP5}.
For $a \approx const.$, $Y$ grows rapidly and we again have to worry about a strong coupling problem as $Y$ approaches 1.
Moreover, the time-derivative of (\ref{equ:last}) implies
\beq
\label{equ:epsM}
2 \epsilon_M \approx -1 - \Bigl( \frac{\d \ln a}{d \ln \tau } \Bigr)^{-1}\ ,
\eeq
where we used that $\epsilon \ll 1$.
For $|\epsilon_M| \ll 1$, (\ref{equ:epsM}) describes perturbative deviations from the near-de Sitter solutions of~\cite{Cheung,SenatoreSmithWMAP5}.
Finally, we consider the limit $|\epsilon_M| \gg 1$, for which (\ref{equ:epsM}) becomes
\beq
\label{equ:epsM2}
2 \epsilon_M \approx  - \Bigl( \frac{\d \ln a}{d \ln \tau } \Bigr)^{-1}\ .
\eeq
We have to worry that the time-dependence of $M(t)$ induces  a strong coupling problem via interactions of the form\footnote{Here we derive the cubic interaction from the decoupled $\pi$-lagrangian~\cite{Cheung} ignoring the mixing with gravity. We expect this to give a lower bound on the level of strong coupling.}
\beq
\label{equ:inte}
\frac{{\cal L}_{\zeta \dot \zeta^2}}{{\cal L}_{\dot \zeta^2}} = \epsilon_M\, \zeta\ . 
\eeq
We see that $\epsilon_M$ cannot be too large. If it starts large and increases with time, we generate large non-Gaussianities. If it decreases with time, it quickly enters the regime $|\epsilon_M| \ll 1$. 
Hence, just as in the case of large $\epsilon$, we impose $\epsilon_M \simeq const$. in order to keep (\ref{equ:inte}) under control, while remaining in the non-inflationary regime $|\epsilon_M| \gg 1$.
In this case we can integrate (\ref{equ:epsM2}) to get
\beq
a \propto \tau^{- 1/2\epsilon_M}\ .
\eeq
However, this power law solution for the scale factor implies $\epsilon = 2 |\epsilon_M| \gg 1$, which is of course inconsistent with our original assumption $\epsilon \ll 1$.

\vskip 4pt
\noindent
{\bf Summary.} \hskip 4pt
Although we haven't given a completely comprehensive analysis of the action (\ref{equ:ghost2}), our analysis of the most promising limiting cases has only revealed the near-de Sitter models of~\cite{Cheung,SenatoreSmithWMAP5} as backgrounds with weakly-coupled scale-invariant fluctuations.

\subsection*{$(\delta E^{i}_{\, i})^2$ Term}

To analyze (\ref{equ:ghost}), it is convenient to define
\beq
X \equiv \frac{M^4}{M_{\rm pl}^2 H^2} \qquad {\rm and} \qquad Y \equiv \frac{\bar M^2}{M_{\rm pl}^2 }\, ,
\eeq
so that the coefficients of the action can be written as
\bea
\bar A &=& (1 + X Y)^{-1} \, (X-9 Y + \epsilon) \ ,\\
\bar B &=&  (1 + X Y)^{-2}\, \Bigl[ -3 Y + XY\big[ XY + \epsilon (1+Y) \big]  +\big[ XY- \epsilon \big] + Y\, \eta \epsilon \Bigr] \ , \\
\bar C &=& -  (1 + X Y)^{-1}\, \frac{Y}{H^2}\ .
\eea
As before, we assume $M \sim \bar M$ and $H/M \ll 1$, so that $|X| \gg |Y|$.
Moreover, we impose $|Y| < 1$ because the cutoff of the effective theory shouldn't exceed the Planck scale.
We notice that the action reduces to the correct limit of slow-roll inflation---$\bar A \to \epsilon$,  $\bar c_s^2 \equiv -\bar B{\bar A}^{-1} \to 1$ and $\bar C\to 0$---for $1 \gg \epsilon \gg |X| > |Y|$.
In order to be maximally different from slow-roll inflation we will restrict our discussion to the limit $|X| \gg \epsilon$:
\bea
\bar A &\to& (1 + X Y)^{-1} \, X \ ,\\
\bar B &\to&  (1 + X Y)^{-2}\, \Bigl[ -3 Y + XY\big[ XY + \epsilon \big]  +\big[ XY- \epsilon \big] + Y\, \eta \epsilon \Bigr] \ , \\
\bar C &\to& -  (1 + X Y)^{-1}\, \frac{Y}{H^2}\ .
\eea
For $Y=0$ this reproduces models with small speed of sound---$\bar A \to X$, $\bar c_s^2 \to \frac{\epsilon}{X} \ll 1$ and $\bar C \to 0$.
For $|XY| \ll 1$, and assuming the higher-derivative term to dominate at horizon crossing (this has to be checked a posteriori for each solution), the action becomes
\beq
\label{equ:ghostXX}
S_2 = M_{\rm pl}^2 \int \d^3 x\, \d \tau\, q^2 \left[(\zeta')^2 - \frac{1}{\mu^2} (\partial_i^2 \zeta)^2 \right]\ ,
\eeq
where
\beq
q^2(\tau) \equiv \frac{a^2 M^4}{M_{\rm pl}^2 H^2} \qquad {\rm and} \qquad \mu^2(\tau) \equiv \frac{a^2 M^4}{\bar M^2}\ .
\eeq
We require a slightly new treatment to determine the conditions under which (\ref{equ:ghostXX}) leads to scale-invariant two-point correlations:
We first consider the equation of motion of the canonically-normalized field $v \equiv \sqrt{2}\, q \zeta$ 
\beq
v_k'' + \left(\omega^2 - \frac{q''}{q} \right) v_k = 0\ , \qquad {\rm where} \qquad  \omega^2(k, \tau) \equiv \frac{k^4}{{\mu}^2} \ .
\eeq
Modes freeze when
\beq
\omega_\star^2 = \left. \frac{q''}{q} \right|_\star \qquad {\rm or} \qquad k^4_\star = {\mu}^2  \left. \frac{q''}{q} \right|_\star\ . 
\eeq
At early times, i.e.~when all modes are deep inside the horizon, the equation of motion reduces to
\beq
v_k'' + \omega^2 v_k = 0 \qquad {\rm or} \qquad  \frac{d}{d\tau} \left[ \frac{1}{2} |v_k'|^2 + \frac{1}{2} \omega^2 |v_k|^2 \right] = \frac{\omega'}{\omega} \, \omega^2 |v_k|^2\ .
\eeq
Averaging over many oscillations, $\langle |v_k'|^2 \rangle = \langle \omega^2 |v_k|^2 \rangle \equiv 2 E$, we find
\beq
\frac{1}{E} \frac{d E}{d \tau} = 
\frac{1}{\omega}\frac{d \omega}{d\tau} \qquad {\rm or} \qquad \frac{E}{\omega} = const.
\eeq
This implies
\beq
\label{equ:ZZeta2}
|\zeta|^2 \propto  \left. \frac{1}{q^2 \omega} \right|_\star = \left. \frac{1}{q^2 ( q''/q)^{1/2}} \right|_\star \ .
\eeq
Assuming $q$ to be a power law in conformal time, 
\beq
q \propto \tau^n\ . \label{equ:taun}
\eeq
this becomes
\beq
\label{equ:ZZeta}
|\zeta|^2 \propto \left. \frac{1}{\tau^{2n -1}} \right|_\star\ , \qquad {\rm where} \qquad
k^4_\star \propto \frac{\mu^2}{\tau^2}  \ .
\eeq
If we parameterize the time-dependence of $\mu$ as a power law, 
\beq
\label{equ:taup}
\mu \propto \tau^{p}\ ,
\eeq
we find $k^4 \propto \tau^{2(p -1)}$ and hence
\beq
|\zeta|^2 \propto k^{-\frac{4n-2}{p-1}} \ .
\eeq
We see that scale-invariance, $|\zeta|^2 \propto k^{-3}$, requires the following algebraic relation between the power law indices
\beq
4 n = 3 p - 1\ .
\eeq
The time-dependence of the canonically-normalized field on superhorizon scales is $v \propto \tau^n$ and $\tau^{-n+1}$.
Hence, the two solutions for
$\zeta \sim v q^{-1}$ scale as $\tau^0$ and $\tau^{-2n+1}$. For $n \le \frac{1}{2}$ the growing mode of $\zeta$ will be a constant and the background is an attractor. We will restrict to that case.

Let us further simplify the treatment by considering solutions with constant $\epsilon$ and hence $a \propto \tau^{1/(\epsilon-1)}$.
The conditions for scale-invariance, $q \propto \tau^n$ and $\mu \propto \tau^{\frac{4n+1}{3}}$, then imply
\beq
\label{family}
 \frac{M^2 \bar M^3}{H^4} \propto \frac{1}{ a\tau} \propto a^{-\epsilon} \qquad \Rightarrow \qquad
2 \epsilon_M + 3 \epsilon_{\bar M} = - 5 \epsilon\ .
\eeq
We have therefore found a whole family of solutions with scale-invariant two-point functions.
This includes {\it ghost inflation}~\cite{GhostInflation}: $\epsilon=0$ and $\epsilon_M = \epsilon_{\bar M} = 0$, i.e.~de Sitter backgrounds, $a = -1/(H \tau)$, with $M =\bar M = const$.
We also find de Sitter backgrounds with possibly large time-variations in the coefficients of the effective action: $\epsilon=0$ and $\epsilon_M = - \frac{3}{2} \epsilon_{\bar M}$.
Finally, the solutions (\ref{family}) contain non-inflationary cases with $\epsilon =const. \gg 1$ and $\{|\epsilon_M|, |\epsilon_{\bar M}| \} = const. \gg 1$.

Let us look at the last solution in a bit more detail.
For $|X| \gg \epsilon = const. \gg 1 \gg |Y|$ and $|XY| \ll 1$, the action (\ref{equ:ghost2}) simplifies considerably
\beq
\label{equ:simpleA}
S_2 = M_{\rm pl}^2 \int \d^3 x \, \d \tau\, q^2 \left[ (\zeta')^2 - \bar c_s^2 (\partial_i \zeta)^2 - \frac{1}{\mu^2} (\partial_i^2 \zeta)^2\right]\ ,
\eeq
where
\beq
\label{coeff}
\bar c_s^2 \to \frac{\epsilon}{X} \ , \qquad
q^2 \to a^2 X = \frac{a^2 \epsilon}{\bar c_s^2}\ ,  \qquad
\mu^2 \to (a H)^2 \,\frac{X}{Y}\ .
\eeq
The $k^4$-term dominates at horizon crossing if
\beq
\label{crossing}
\bar c_s^2 k^2_\star \ll \frac{k^4_\star}{\mu^2}\ , 
 \eeq
where
 \beq
 \label{kstar}
k_\star^4 = \mu^2 \frac{q''}{q} = n(n-1) \frac{\mu^2}{\tau^2} = n(n-1) \, \epsilon^2\, (aH)^4 \frac{|X|}{|Y|}\ .
 \eeq
 Substituting (\ref{coeff}) and (\ref{kstar}) into (\ref{crossing}) we find
 \beq
 1 \ \gg \ |XY| \ \gg\ \frac{1}{n(n-1)}\ .
 \eeq
Since we restricted to $n < \frac{1}{2}$ for attractor solutions, this implies
\beq
\label{upper}
n \ll \frac{1}{2}(1 - \sqrt{5}) \approx - 0.6\ .
\eeq
This forces the coefficients of the action to be strongly time-dependent.
We should therefore check the time-dependence of interactions.
Consider, for example,
\beq
{\cal L}_{\zeta' (\partial_i \zeta)^2} = M_{\rm pl}^2 q^2 \frac{\zeta'}{aH} (\partial_i \zeta)^2\ ,
\eeq
and compare it to
\beq
{\cal L}_{(\zeta')^2} = M_{\rm pl}^2 q^2 \, (\zeta')^2\ .
\eeq
Using $\omega = k^2/\mu$, we find
\beq
\frac{{\cal L}_{\zeta' (\partial_i \zeta)^2} }{{\cal L}_{(\zeta')^2} } = \frac{\mu}{aH} \zeta = \frac{|X|^{1/2}}{|Y|^{1/2}} \, \zeta\ ,
\eeq
where
\beq
\frac{|X|}{|Y|} =  \frac{|X|_i}{|Y|_i}  \, \Big( \frac{\tau}{\tau_i}\Big)^{\frac{8}{3}(n+1)}\ .
\eeq
Here, we have used that $\mu \propto \tau^{\frac{4n+1}{3}}$ and $aH \propto \tau^{-1}$.
Since $|X|_i \gg |Y|_i$, this is a large interaction at some fiducial time $\tau_i$. Moreover, it grows, unless $n \ge -1$.
Together with (\ref{upper}), this essentially rules out the scenario. This case is therefore another example where two-point correlations are perfectly scale-invariant, but higher-order correlations are forced to have a strong time-dependence.
The fluctuations therefore again can't be weakly coupled over a sufficiently large range of scales.

\vskip 4pt
Finally, we consider the opposite limit $|XY| \gg \epsilon \approx const. > 1$.
In this case the action is of the same form as (\ref{equ:simpleA}), but the coefficients now being
\beq
\label{coeff2}
\bar c_s^2 \to -Y \ , \qquad
q^2 \to \frac{a^2}{Y} =-  \frac{a^2}{\bar c_s^2}\ ,  \qquad
\mu^2 \to (a H)^2 \,\frac{X}{Y}\ .
\eeq
The $k^4$-term dominates at horizon crossing if
\beq
\epsilon \ \ll\ |XY| \ \ll\ \epsilon^2 |n(n-1)|\ .
\eeq
For $\epsilon > 1$ this is easily satisfied. 
Note that our previous conditions for scale-invariance, $q \propto \tau^n$ and $\mu \propto \tau^{\frac{4n+1}{3}}$, still apply.
This time an interaction proportional to $\dot {\bar M}$ creates the most obvious problems:
\beq
\frac{{\cal L}_{\zeta (\partial_i^2 \zeta)^2} }{{\cal L}_{(\zeta')^2}} = \epsilon_{\bar M} \, XY\, \zeta\ ,
\eeq
where
\beq
|XY| = |XY|_i \, \Big( \frac{\tau}{\tau_i}\Big)^{\frac{4}{3}(2-n)}\ .
\eeq
In order for this large interaction to be time-independent we require $n \approx 2$ (for $n > 2$ the theory quickly becomes strongly coupled, whereas for $n < 2$ we quickly exit the regime of interest $|XY| \gg \epsilon$.). However, this is larger than our upper bound on attractor solutions $n < \frac{1}{2}$. We therefore don't find a successful attractor solution in the regime $|XY| \gg \epsilon$.

\vskip 4pt
\noindent
{\bf Summary.} \hskip 4pt We have reproduced ghost inflation, but found no successful non-de Sitter backgrounds with scale-invariant weakly-coupled fluctuations.

\newpage
\section{Comments on Non-Attractor Solutions}
\label{sec:nonA}

In this appendix we consider the second type of backgrounds that lead to scale-invariant fluctuations: 
\beq
\label{equ:CC2}
q^2 = \frac{a^2 \epsilon}{c_s} = q_i^2\, \frac{y^4}{y_i^4}\ .
\eeq
Since the curvature perturbations $\zeta$ in this case evolve outside of the horizon as $\zeta \propto y^{-3}$, we can't make definitive predictions for observations without specifying the evolution of $\zeta$ after the scaling phase. 
We will work throughout with the simplifying assumption that $\zeta$ freezes immediately after the scale-invariant phase ends. The history of the FRW background during the scaling regime is then sufficient to predict the primordial power spectrum relevant for observations.
Many of these results have previously been derived by Khoury and Piazza~\cite{Justin}. 
This section is a concise summary of their analysis with a few corrections and clarifications.

\subsection{Basic Challenges for a Predictive Theory}

The non-attractor nature of the background implies that additional physics has to be added before and after the scaling phase.
This significantly complicates even the minimal scenario.
We will first indicate the basic model-building challenges, before adding more technical details and an analysis of the strong coupling problem in the following sections.

\vskip 4pt
The basic timeline of the universe is:
\begin{itemize}
\item {\it Pre-scaling phase}

Since inhomogeneities grow exponentially during the scaling regime they need to be exponentially small at the beginning of the scaling regime.
To avoid an enormous fine-tuning of the initial conditions requires a pre-scaling phase with the right physical characteristics to set the initial conditions dynamically, i.e.~the universe needs to be cleaned from pre-existing inhomogeneities to extremely high precision. This is similar to the cleaning mechanism proposed for New Ekpyrotic Cosmology~\cite{Buchbinder:2007tw}.

\item {\it Scaling phase}

The background then develops a specific time-evolution of the speed of sound that allows for the generation of scale-invariant curvature perturbations.
If $a(y)$ and $c_s(y)$ are power law solutions, scale-invariance can only be achieved if $c_s$ changes rapidly with $y$. This is consistent with the result that curvature perturbations in a contracting spacetime have a very blue spectrum if $c_s=1$~\cite{CreminelliAttractor}. Just as in the case of the exponentially changing $\epsilon$ parameter above, the  exponential change of $c_s$ is reason to worry about the strong coupling problem.
However, now there are two independent functions of time -- $a(y)$ and $c_s(y)$ -- as well as the time-dependent curvature perturbation $\zeta(y)$, so we expect more freedom to evade constraints.

\item {\it Post-scaling attractor}

A crucial challenge for the non-attractor cases is to make contact with late-time cosmological observables. 
The time-evolution of curvature fluctuations needs to be followed even after the modes exit the horizon. During the scaling phase this time-evolution is fixed, however, after the scaling phase there is some freedom in matching to an attractor solution that connects continuously to the FRW expansion after the bounce.
To avoid introducing an element of unpredictivity, we follow Khoury and Piazza~\cite{Justin} in assuming that the background transitions instantaneously from the scaling phase to a post-scaling attractor, i.e.~$\zeta$ (and $c_s$) freeze immediately after the scaling phase. Observables then only depend on the physics during the scaling regime.

\item {\it Bounce}

As usual in contracting models, the background has to bounce to connect to the expanding FRW phase.
If the background has approached the post-scaling attractor before the bounce then the scale-invariance in the pre-bounce fluctuations will be preserved in the post-bounce fluctuations~\cite{CreminelliAttractor} (unless the unknown physics of the bounce is sensitive to exponentially small decaying modes).

\end{itemize}

Given these minimal model-building requirements, we next discuss the basic constraints arising from scale-invariance and weak coupling.

\subsection{Background Solution}
\label{sec:scaling}

To characterize the time-dependence of the background during the scaling regime, 
we define the following two parameters
\beq
\epsilon = - \frac{\dot H}{H^2} \qquad {\rm and} \qquad \epsilon_s = \frac{\dot c_s}{H c_s}\ .
\eeq
In terms of derivatives with respect to $y$ ($d/dy='$) these can be expressed as
\beq
\epsilon_s = \frac{(\ln c_s)'}{(\ln a)'}  \qquad {\rm and} \qquad 
2 - \epsilon_s - \epsilon = \frac{a''/a}{(a'/a)^2}\, . 
\eeq
Assuming that $\epsilon$ and $\epsilon_s$ are constant gives power law solutions~\cite{Justin}
\beq
\label{equ:evo}
a(y) \sim (-y)^{p} \quad {\rm and} \quad c_s(y) \sim (- y)^{p \, \epsilon_s }\, , \quad {\rm with} \quad p \equiv \frac{1}{\epsilon_s + \epsilon - 1}\ ,
\eeq
and
\beq
\label{equ:qq2}
q^2 \sim (-y)^\delta \, , \quad {\rm with} \quad \delta \equiv \frac{2-\epsilon_s}{\epsilon_s + \epsilon -1}\ .
\eeq
Comparing (\ref{equ:qq2}) to (\ref{equ:CC2}) we find
\beq
{\sf Case \ II} \quad \Rightarrow \quad  \delta = 4   \quad \Rightarrow \quad  \fbox{$\displaystyle \epsilon_s = \frac{2}{5} (3-2\epsilon)$}  \ ,\label{Case2b}
\eeq
and
\beq
\label{equ:back}
a(y) \sim (-y)^{\frac{5}{1+\epsilon}} \quad {\rm and} \quad c_s(y) \sim (-y)^{\frac{6-4\epsilon}{1+\epsilon}}\ .
\eeq
Via the algebraic relation between $\epsilon$ and $\epsilon_s$ in (\ref{Case2b}) the combined time-dependence of $a$ and $c_s$ ensures scale-invariance of the two-point function.
In the analysis of the strong coupling problem the time-dependence of the term $\zeta/c_s^2$ between horizon exit and the end of the scaling phase will be essential,
 \beq
 \label{equ:coupling}
\fbox{$\displaystyle \frac{\zeta}{c_s^2} =\frac{\zeta_{\rm end}}{c_{s, \rm end}^2}\left(\frac{y}{y_{\rm end}}\right)^\beta $} \qquad {\rm with} \qquad \fbox{$\displaystyle \beta \equiv \frac{5(\epsilon-3)}{1+ \epsilon} $}\ ,
\eeq
where $\zeta_{\rm end}$ and $c_{s,\rm end}$ are constrained by observations: $\zeta_{\rm end} \sim 10^{-5}$ and $c_{s,\rm end}\gtrsim 0.01$.\\

\noindent
{\small {\sl \underline{Digression}: spacetime during the scaling phase}
\vskip 4pt

Since inhomogeneities grow during the non-attractor phase it is interesting to determine what the spacetime would look like.
To first order in $\zeta$, we find the following metric~\cite{CreminelliAttractor}
\beq
\d s^2 = - \Bigl[ 1 +2 \frac{\dot \zeta}{H} \Bigr] \d t^2 +   \partial_i \Bigl[ \frac{a^2 \epsilon}{c_s^2} \frac{1}{\nabla^2} \dot \zeta - \frac{\zeta}{H}\Bigr]\, \d t \d x^i +  a^2 (1+2 \zeta) \delta_{ij}\, \d x^i \d x^j\ .
\eeq
The intrinsic curvature is 
\beq
\frac{{}^{(3)}R}{H^2} = - 4 \frac{\nabla^2 \zeta}{(aH)^2}\ .
\eeq
In the background (\ref{equ:back}) this has the following time-dependence
\beq
\frac{{}^{(3)}R}{H^2} \sim (k y)^2 |y|^\beta\ .
\eeq
The intrinsic curvature perturbation is hence suppressed on superhorizon scales $|k y| \ll 1$.
Furthermore, unless $\beta$ is very negative it decreases with time.
From the metric we can also compute the extrinsic curvature
\beq
K^i_{\ j} = \frac{1}{2} g^{ik} \left( \dot g_{kj} - \partial_k g_{0j} - \partial_j g_{k0} \right)\ .
\eeq
To first order in $\zeta$, we find 
\beq
\frac{\delta K^i_{\ j}}{H} = \frac{\dot \zeta}{H} \delta^i_j - \frac{\partial_i \partial_j }{\nabla^2} \left[ \frac{\epsilon}{c_s^2} \frac{\dot \zeta}{H}+ \frac{1}{4}\frac{{}^{(3)}R}{H^2}  \right]\ ,
\eeq
where $\bar K^i_{\ j} = H \delta^i_j$ is the extrinsic curvature of the background.
Since $ \dot \zeta H^{-1}$ has the same time-dependence as $\zeta$,
\beq
\frac{\dot \zeta}{H} = \frac{\zeta'}{(\ln a)'} \sim |y|^{-3}\ ,
\eeq
the isotropic component of the extrinsic curvature tensor grows exponentially relative to the background.
The time-dependence of the anisotropic component of the extrinsic curvature follows from
\beq
\frac{1}{c_s^2} \frac{\dot \zeta}{H} \sim |y|^\beta\ .
\eeq
Only for $\beta$ not too negative will the anisotropic extrinsic curvature be under control.}

\subsection{Strong Coupling Constraints}

When is the time-evolution of $a$ and $c_s$ consistent with weak coupling over a large range of scales?

As before, we require that the theory is weakly coupled at horizon crossing, i.e.
\beq
X \equiv \left. \frac{{\cal L}_{3}}{{\cal L}_{2}} \right|_\star \sim \left. {\cal O}(\{1, \epsilon, \eta, \epsilon_s \}) \, \frac{\zeta}{c_s^2}\right|_\star\ \ll\ 1\ .
\eeq
This guarantees that {\it quantum mechanical} loop effects are small and the perturbative computation of the primordial fluctuations is under control.
Furthermore, the growth of $\zeta$ on superhorizon scales can lead to the {\it classical} growth of non-linearities. Keeping these non-Gaussianities below current observational constraints imposes further constraints on the parameter space.

\vskip 6pt
\noindent
{\bf Quantum mechanical constraints.} \hskip 4pt
\label{sec:quantum}
Eqn.~(\ref{equ:coupling}) describes the size of ${\cal L}_3/{\cal L}_2$ at horizon crossing. We distinguish the three cases $\beta = 0$, $\beta < 0$  and $\beta > 0$:

\vspace{-0.2cm}
\subsubsection*{$\beta = 0$} 

For $\epsilon = 3$ ({i.e.}~pure kinetic-dominated contraction) the time-dependence of the speed of sound, $c_s^2 \propto y^{-3}$, exactly compensates for the time-dependence of $\zeta$ -- i.e.~both the speed of sound and the curvature perturbation are exponentially small at the time when the first modes exit the horizon:\footnote{The model therefore features a couple of very small (but technically natural) numbers: e.g.~the sound speed at the beginning of the scaling phase is $c_s^2(y_i) = 10^{-13}$ for $\Delta N = 10$ and $c_s^2(y_i) = 10^{-65} $ for $\Delta N = 50$. Furthermore, using $H^2c_s^{-1} \sim y^{-6}$ we  find $H(y_i) \sim e^{-10 -\frac{15}{4} \Delta N} M_{\rm pl}$, i.e.~the Hubble parameter at the beginning of the scaling regime is $H(y_i) = 10^{-21} M_{\rm pl}$ for $\Delta N = 10$ and $H(y_i) = 10^{-86} M_{\rm pl}$ for $\Delta N = 50$.} 
$c_s^2(y_i) < (e^{\Delta N})^{-3} \sim e^{-3 \Delta N}$ and $\zeta \sim \zeta_{\rm end}\, e^{-3\Delta N}$, where $\Delta N$ denotes the number of $e$-folds of contraction between $y_i$ and $y_{\rm end}= e^{- \Delta N} y_i$. 
The time-independence of $\zeta/c_s^2 \sim 10^{-5}$ for $\beta = 0$ implies the absence of the quantum mechanical strong coupling problem. 
However, below we show that the superhorizon generation of non-Gaussianities in this case leads to a classical strong coupling problem that rules out the $\beta \approx 0$ limit of the parameter space.

\subsubsection*{$\beta < 0$} 

For $\epsilon < 3$ the fluctuations are more weakly coupled at the initial time $y_i$ than at the final time $y_{\rm end}$.
Hence, if the weak coupling condition is satisfied for the mode that exits the horizon at $y_{\rm end}$ (as required by observations), then it will automatically be satisfied by all modes that exit the horizon at earlier times $y < y_{\rm end}$.

\subsubsection*{$\beta > 0$} 

For $\epsilon > 3$ the fluctuations are more strongly coupled at the initial time $y_i$ than at the final time $y_{\rm end}$.
This fact implies an upper limit on $\epsilon$ arising from the fact that we want to avoid a strong coupling problem for the largest CMB scales, which cross the horizon at $y_i$.
Quantitatively, the constraint
\beq
X(y_i)= {\cal O}(1) \cdot \frac{\zeta_{\rm end}}{c_{s, \rm end}^2}\, e^{\beta \Delta N} \ \ll \ 1\ ,
\eeq
implies
\beq
\beta  \ \lesssim  \ \frac{10}{\Delta N} \qquad \Rightarrow \qquad  \epsilon \ \lesssim \ 3 + \frac{8}{\Delta N} \ .  \label{equ:upper}
\eeq
Observations imply at least $\Delta N \sim 10$ $e$-folds of scale-invariant fluctuations. The constraint in (\ref{equ:upper}) then becomes $\epsilon \lesssim 3.8$.
However, below we will see that for any value of $\beta > 0$ ($\epsilon > 3$) the non-Gaussianities developed by superhorizon evolution quickly lead to a strong coupling problem.

\vskip 6pt
\noindent
{\bf Classical constraints.} \hskip 4pt
\label{sec:classical}
The non-linear growth of $\zeta$ on superhorizon scales leads to an additional accumulation of non-Gaussianities.
In this subsection we identify the regime of parameter space for which those non-Gaussianities remain small enough to be consistent with observations.
In Appendix~\ref{sec:bispectra} we compute the exact bispectra for the background of \S\ref{sec:scaling}, following Khoury and Piazza~\cite{Justin}:
\beq
\langle \zeta_{{\bf k}_1}  \zeta_{{\bf k}_2}  \zeta_{{\bf k}_3}\rangle = (2\pi)^7\, \delta({\bf k}_1 + {\bf k}_2 + {\bf k}_3)\, \frac{P_\zeta^2}{\prod_i k_i^3} \, {\cal A}\ .
\eeq
We will use two different measures for the amount of non-Gaussianity:
\beq
f_{\rm NL} \equiv \frac{40}{3} \frac{1}{K^3} \ \lim_{k_1 \to 0} {\cal A}\, ,
\eeq
for non-Gaussianity in the squeezed limit, and
\beq
\widehat f_{\rm NL} \equiv 30\, \frac{1}{K^3}\  {\cal A}_{k_1=k_2=k_3}\, ,
\eeq
for non-Gaussianity of general shape.
Here, we have defined
$K \equiv k_1 + k_2+k_3$.
As before, we will describe our results in term of deviations from the sweet spot $\epsilon = 3$, or $\beta = 0$.
The details are presented in Appendix~\ref{sec:bispectra}. Here we summarize our findings:

\subsubsection*{$\beta = 0$} 

In the limit $\beta = 0$ simple analytical expressions for the non-Gaussianities can be obtained.
For instance, the interaction $\dot \zeta^3$ produces the following non-Gaussianity in the squeezed limit
\bea
\label{equ:fNLmax}
\lim_{\beta \to 0} f_{\rm NL}^{\dot \zeta^3} &=&
 \frac{24}{c_{s\, \rm end}^2}  ( f_X^{(0)}-1) \cdot \log (k/k_{\rm end})  
\ +\ \cdots\ ,
\eea
where $f_X^{(0)}$ is an order unity coefficient related to coefficient of the $(\delta g^{00})^3$ term in the effective Lagrangian of \cite{Cheung} (see Appendix~\ref{sec:bispectra}).
Unless one ``tunes"\footnote{This ``tuning" may have a microphysical justification. In fact, curiously, if the background arises from a Dirac-Born-Infeld (DBI) action then $f_X^{(0)} = 1$ and the contribution in (\ref{equ:fNLmax}) vanishes.} $f_X^{(0)}$ this gives a large scale-dependent non-Gaussianity. This limits the scaling regime to a few $e$-folds.
Specifically, for $c_{s, \rm end} \sim 1$, $|f_X^{(0)} -1| \sim {\cal O}(1)$, and $k/k_{\rm end} = e^{-\Delta N}$, Eqn.~(\ref{equ:fNLmax}) becomes
\beq
\lim_{\beta \to 0} |f_{\rm NL}^{\dot \zeta^3}|  \approx 24\, \Delta N\ .
\eeq
The observational limit $|f_{\rm NL}| \lesssim 100$ then implies
\beq
\Delta N \lesssim 5\ .
\eeq
This is of course inconsistent  with the observationally required range of scale-invariant modes, $\Delta N \gtrsim 10$.
This result is not a quantum disaster as before (in the sense of $\frac{{\cal L}_3}{{\cal L}_2} > 1$ at horizon crossing), but only an observational fact about our universe.
Moreover, 
we should remark that this constraint can be avoided if $|f_X^{(0)} - 1| \ll 1$.
In fact, if the background is sourced by a Dirac-Born-Infeld (DBI) action then $f_X^{(0)} = 1$ and the dangerous contribution in (\ref{equ:fNLmax}) vanishes.
In that case, the leading non-Gaussianity comes from two non-local interactions (see Appendix~\ref{sec:bispectra}).
Individually each of these interactions produces a large scale-dependent non-Gaussianity.
However, in the squeezed limit the two terms cancel precisely up to terms proportional to $\beta$,
\beq
\lim_{\beta \to 0} \ ( f_{\rm NL}^{\, \dot \zeta \partial \zeta \partial \chi} + f_{\rm NL}^{\, \epsilon^2} ) = {\cal O}(\beta)\ .
\eeq
 Remarkably, for $\beta=0$ the combined non-Gaussianity produced by these interactions is therefore identically zero in the squeezed limit.
However, this cancelation doesn't hold in the equilateral configuration, as we see from the following result:
\bea
\lim_{\beta \to 0} \ ( \widehat f_{\rm NL}^{\ \dot \zeta \partial \zeta \partial \chi} + \widehat f_{\rm NL}^{\,\epsilon^2} ) &=&  - \frac{1}{c_{s\, \rm end}^2} \frac{375}{32} \log(k/k_{\rm end}) + \frac{2.3}{c_{s\, \rm end}^2}   \ \approx \ 10\, \Delta N\ .
\eea
At best this is marginally consistent with the observational limit $\widehat f_{\rm NL} \lesssim 300$~\cite{SenatoreSmithWMAP5}.

The above proves that the special case $\beta = 0$ (corresponding to pure kinetic energy domination), although consistent with ${\cal L}_3 \ll {\cal L}_2$ at horizon crossing, is ruled out by the size of the non-Gaussianity arising from the superhorizon evolution of $\zeta$ (unless $f_X^{(0)}$ is close to 1.2---see below).

 \begin{figure}[h!]
    \centering
        \includegraphics[width=.7\textwidth]{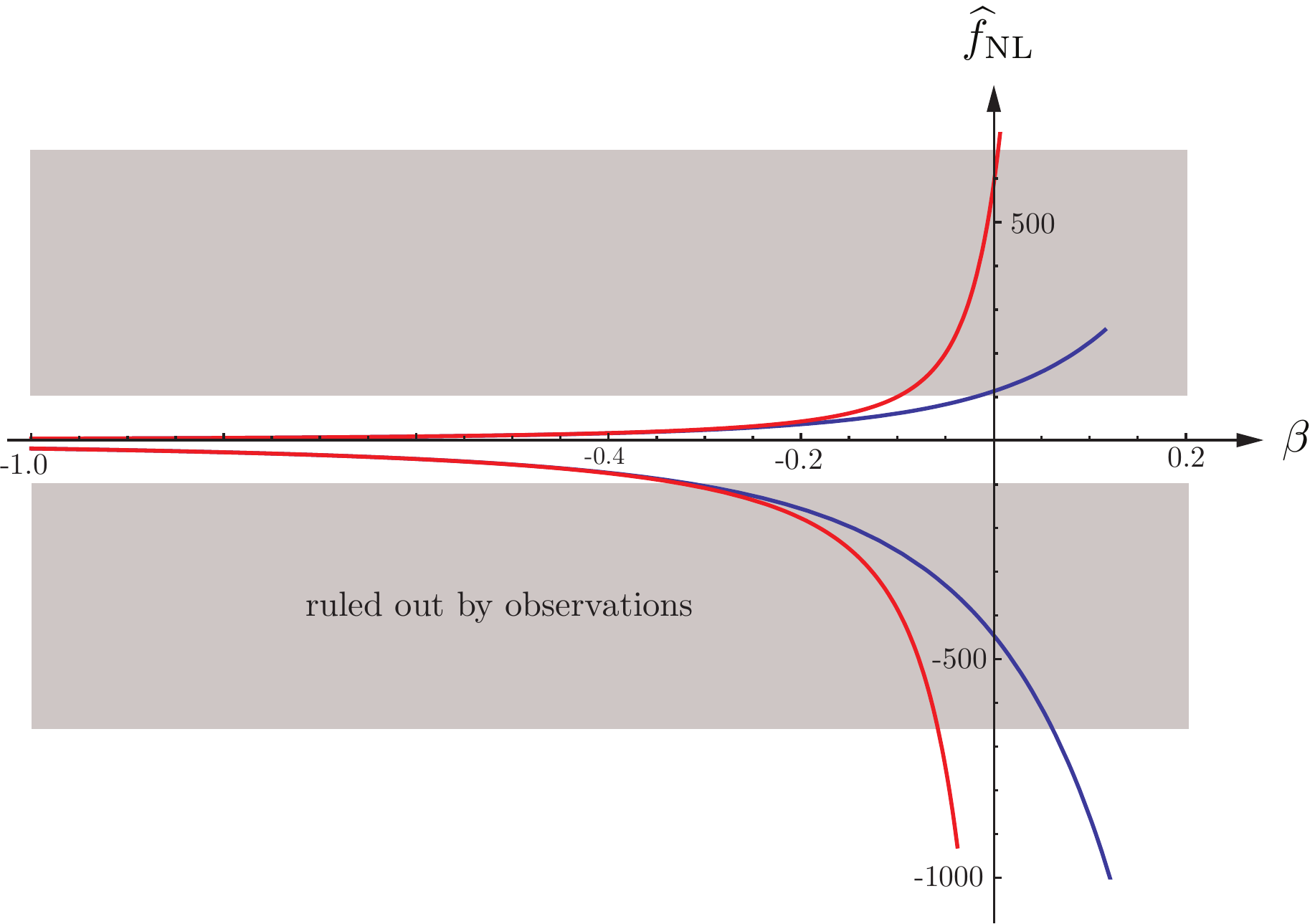}
    \caption{\sl Plot of $\widehat f_{\rm NL}$ vs. $\beta$ for $y_i = e^{-10} y_{\rm end}$ (blue) and $y_i = e^{-50} y_{\rm end}$ (red). We show two cases: $f_X^{(0)} = 1$ (DBI) ($\Rightarrow \widehat f_{\rm NL} > 0$) and  $(f_X^{(0)} - 1)= 1$ ($\Rightarrow \widehat f_{\rm NL} < 0$). The plot implies an upper limit on $\beta$ but {\it no} lower limit. This difference from the result of \cite{Justin} arises from corrections to their result for the bispectrum (see Appendix~\ref{sec:bispectra}).}
    \label{fig:fNL}
\end{figure}

\subsubsection*{$\beta < 0$} 

For general $\beta$ we present the result graphically in Figure~\ref{fig:fNL} (for specific values of $f_X^{(0)}$). The plot confirms that the magnitude of $\widehat f_{\rm NL}$ is inconsistent with observational limits for $\beta \ge 0$. However, it
also shows that $\widehat f_{\rm NL}(\beta)$ is strongly suppressed for negative $\beta$ (or $\epsilon < 3$).
For generic $f_X^{(0)}$---i.e.~$|f_X^{(0)}-1|=1$---we find a new upper limit for the equation of state during the scaling regime:
\beq
\label{equ:bound}
\beta < - 0.3 \qquad \Rightarrow \qquad \epsilon < 2.77\ .
\eeq
For $\epsilon < 2.77$ the non-Gaussianity is still consistent with present observational constraints.
Moreover, if $f_X^{(0)}$ is tuned to lie between 1.1 and 1.3, the non-Gaussianity is sufficiently small for all negative $\beta$.
 
 It is interesting to compare this upper bound on $\epsilon$ with a lower bound obtained by requiring that the background during the scaling phase solves the flatness, horizon and isotropy problems of standard big bang cosmology.
 Consider the Friedmann equation
 \beq
 H^2 = - \frac{k}{a^2} + \frac{8}{3M_{\rm pl}^2} \left( \frac{c_m}{a^3} + \frac{c_\gamma}{a^4} + \frac{c_{\rm a}}{a^6} \right) + \frac{8}{3M_{\rm pl}^2}  \, \frac{c_X}{a^{2\epsilon}}\ ,
 \eeq
 where we have allowed for contributions from curvature, matter, radiation and anisotropies, in addition to the fluid $X$ which sources the background during the scaling regime.
We see that $\epsilon > 3$ is required such that pre-existing anisotropies don't come to dominate the universe.
 Interestingly, this is inconsistent with the bound in (\ref{equ:bound}).
 This implies that in order to make the scenario viable, a phase has to be added before the scaling regime which reduces anisotropies to such a low level that their subsequent growth remains harmless.
 This is to be contrasted with {\sf Case I} (inflation) where backgrounds with $\epsilon < 1$ automatically solve the flatness problem without added features.

\section{Bispectra for Non-Attractor Solutions}
\label{sec:bispectra}

In this appendix we compute the complete bispectra of curvature fluctuations in the non-attractor solutions of Appendix~\ref{sec:nonA}, following the methodology developed in \cite{Justin}.
Our starting point is the cubic action for $\zeta$:\footnote{This action follows from the effective Lagrangian of Ref.~\cite{Cheung} by including the effects of mixing with metric perturbations~\cite{Chen}. The order-one parameter $f_X$ is model-dependent---in the effective theory of inflation it arises from the coupling of the $(\delta g^{00})^3$ term of the action---(see \cite{Cheung, Justin} for the precise definition). If the action (\ref{action3}) is derived from the DBI action then $f_{X}^{\rm DBI} = (1-c_s^2)$~\cite{Justin} or $\tilde c_3^{\rm DBI} = \frac{3}{2}(1-c_s^2)$~\cite{SenatoreSmithWMAP5}. Here and in the following we have set $M_{\rm pl} \equiv 1$.}
\begin{eqnarray} \label{action3}
S_{\rm int}&=&\int {\rm d}t \,{\rm d}^3x\, \left\{ -
\frac{2}{3} \frac{H^2 \epsilon}{c_s^4} \Bigl[ f_X - (1-c_s^2)\Bigr] \frac{a^3}{H^3} \, \dot{\zeta}^3
+\frac{a^3\epsilon}{c_s^4}(\epsilon-3+3c_s^2)\, \zeta\dot{\zeta}^2 \right.
\nonumber \\ &+&
\frac{a\epsilon}{c_s^2}(\epsilon-2\es+1-c_s^2)\, \zeta(\partial_i\zeta)^2-
2a \frac{\epsilon}{c_s^2}\, \dot{\zeta}(\partial_i
\zeta)(\partial_i \chi) \nonumber \\ &+& \left.
\frac{a^3\epsilon}{2c_s^2}\frac{d}{dt}\left(\frac{\eta}{c_s^2}\right)\zeta^2\dot{\zeta}
+\frac{\epsilon}{2a}(\partial_i\zeta)(\partial_i
\chi) \partial^2 \chi +\frac{\epsilon}{4a}(\partial^2\zeta)(\partial_i
\chi)^2+ 2 f(\zeta)\left.\frac{\delta {\cal L}_{2}}{\delta \zeta}\right\vert_1 \right\} ~,
\end{eqnarray}
where dots denote derivatives with respect to proper time $t$, $\partial_i$ is a spatial derivative,
and $\chi$ is defined as
\begin{equation}
\partial^2 \chi \equiv \frac{a^2 \epsilon}{c_s^2}\dot{\zeta}\,.
\end{equation}
The term $\frac{\delta {\cal L}_{2}}{\delta\zeta}|_1$ denotes the variation of the
quadratic action with respect to the perturbation $\zeta$:
\beq
\left.\frac{\delta
{\cal L}_{2}}{\delta\zeta}\right\vert_1 = a
\left( \frac{d\partial^2\chi}{dt}+H\partial^2\chi
-\epsilon\partial^2\zeta \right) \ .
\eeq
Finally, we have defined the function
\begin{eqnarray} \label{redefinition}
f(\zeta)&\equiv&\frac{\eta}{4c_s^2}\zeta^2+\frac{1}{c_s^2H}\zeta\dot{\zeta}+
\frac{1}{4a^2H^2}[-(\partial_i \zeta)(\partial_i \zeta)+\partial^{-2}(\partial_i\partial_j(\partial_i\zeta\partial_j\zeta))] \nonumber \\
&+&
\frac{1}{2a^2H}[(\partial_i\zeta)(\partial_i\chi)-\partial^{-2}(\partial_i\partial_j(\partial_i\zeta\partial_j\chi))] ~,
\end{eqnarray}
where $\partial^{-2}$ is the inverse Laplacian.

In an impressive tour the force Khoury and Piazza~\cite{Justin} computed the bispectra for the action (\ref{action3}). 
In this appendix we reproduce their results (with small corrections) for the case of constant $\epsilon$, $\epsilon_s$ and $\tilde c_3$.
We will use the standard expression for the three-point function in the in-in formalism:
\beq
\langle \hat \zeta_{{\bf k}_1} \hat \zeta_{{\bf k}_2}  \hat \zeta_{{\bf k}_3} \rangle = - i \int_{-\infty}^0 \d y'\, \langle [  \hat \zeta(0,{\bf k}_1) \hat \zeta(0,{\bf k}_2)  \hat \zeta(0,{\bf k}_3) , \hat H_{\rm int}(y')]  \rangle\ .
\eeq
The operators $\hat \zeta$ are expanded in creation and annihilation operators in the usual way:
\beq
\hat \zeta(y, {\bf k}) =  \zeta_k(y) \hat a_{\bf k} + \zeta_k^*(y) a^\dagger_{-{\bf k}}\ , 
\eeq
where $[ \hat a_{\bf k}, \hat a^\dagger_{-{\bf k}'}] = (2\pi)^3 \delta({\bf k}+{\bf k}')$ and the mode functions are~\cite{Justin}
\beq
\zeta_k(y) = \frac{H(1-\epsilon - \epsilon_s)}{\sqrt{4c_s \epsilon k^3}} (1+iky) e^{-iky}\ .
\eeq
Finally, we recall that the background satisfies
\beq
\frac{H}{c_s^{1/2}} = \frac{H_{\rm end}}{c_{s\; {\rm end}}^{1/2}} \left( \frac{y}{y_{\rm end}}\right)^{-3} \qquad {\rm and} \qquad \frac{H}{c_s^{1/2}} = \frac{H_{\rm end}}{c_{s\; {\rm end}}^{1/2}} \left( \frac{y}{y_{\rm end}}\right)^{\beta} \ ,
\eeq
where $\beta \equiv 5(\epsilon-3)(\epsilon+1)^{-1}$.

\subsection{Shapes of Non-Gaussianity}

With these ingredients we compute the bispectra associated with the individual interactions in (\ref{action3}).
Defining $K \equiv k_1 + k_2 + k_3$ and
\beq
\langle \hat \zeta_{{\bf k}_1} \hat \zeta_{{\bf k}_2}  \hat \zeta_{{\bf k}_3} \rangle = (2\pi)^7 \delta({\bf k}_1 +{\bf k}_2+{\bf k}_3) \frac{P_\zeta^2}{\prod_i k_i^3} \, {\cal A}\ ,
\eeq
we find:
\bea
\nonumber
{\cal A}^{(\bf II)}_{\dot\zeta^3} &=& \frac{(1+\epsilon)}{10\, c_{s\;{\rm end}}^{2}} \left(f_X^{(0)} - 1 \right)\left\{\frac{\cos\frac{\pi\beta}{2}}{(K|y_{\rm end}|)^\beta}\left[\Gamma(\beta+3)\frac{k_1^2k_2^2k_3^2}{K^3} \nonumber
+ 3\Gamma(\beta+2)\frac{k_1k_2k_3}{K^2}\sum_{i<j}k_ik_j \right.\right. \\
\nonumber
&+ & \left.\left. 3\Gamma(\beta+1) \left(\frac{1}{K} \sum_{i<j} k_i^2 k_j^2 + 3 k_1 k_2 k_3\right)+  9\frac{\Gamma(\beta+1)}{\beta-1}K \sum_{i<j} k_i k_j \right] \right. \\ \nonumber
&+ & \left. \frac{9K^3}{(\beta-2)(\beta-3)}\left[\frac{\cos\frac{\pi\beta}{2}}{(K|y_{\rm end}|)^\beta}\Gamma(\beta)\frac{\beta^2-2\beta+3}{\beta-1} + \frac{3}{\beta}\right]  \right\}
 \, ,\nonumber 
\eea
where we defined $f_X \equiv f_X^{(0)} + f_X^{(2)} c_s^2$,
\bea
\nonumber
{\cal A}^{(\bf II)}_{\zeta\dot\zeta^2} &=& \frac{(\epsilon-3)}{4\, c_{s\;{\rm end}}^{2}}\left\{\frac{\cos\frac{\pi\beta}{2}}{(K|y_{\rm end}|)^\beta}
\left[\Gamma(\beta+1)\left(\frac{2+\beta}{K}\sum_{i<j}k_i^2k_j^2 -\frac{1+\beta}{K^2}\sum_{i\neq j}k_i^2k_j^3+6k_1k_2k_3\right) \right.\right. \\
\nonumber
&+ & \left.\left. 3\Gamma(\beta) \left(2\sum_{i\neq j}k_ik_j^2+9k_1k_2k_3 + \frac{9K^3}{(\beta-2)(\beta-3)}\right) \right. \right. \\
\nonumber
&+& \left.\left. 3\Gamma(\beta-1)K\left(2K^2+5\sum_{i<j} k_ik_j\right) \right] \right. \nonumber \\
\nonumber
&+ & \left. \frac{3}{\beta}\left(2K\sum_{i<j}k_ik_j +\sum_{i\neq j}k_ik_j^2 - K^3\frac{\beta^2-5\beta-3}{(\beta-2)(\beta-3)}\right) \right\} \\
&+& \frac{3}{4} \sum_i k_i^3 
\nonumber\ ,
\eea
and
\bea
\nonumber
{\cal A}^{(\bf II)}_{\zeta(\vec{\nabla}\zeta)^2} &=& \frac{ 13\, \epsilon-7}{40\, c_{s\;{\rm end}}^{2}}\frac{\cos\frac{\pi\beta}{2}}{(K|y_{\rm end}|)^\beta}
\Gamma(\beta+1) \left(\sum_i k_i^2\right)  \cdot \left\{\frac{K}{\beta-1} + \frac{1}{K}\sum_{i<j}k_ik_j +(1+\beta)\frac{k_1k_2k_3}{K^2}\right\}\ .
\eea
The following two terms are from the non-local interactions in $S_{\rm int}$:
\bea
\nonumber
{\cal A}^{(\bf II)}_{\dot \zeta \partial \zeta \partial \chi} &=& - \frac{\epsilon}{4\, c_{s\ \rm end}^2 }\left\{\frac{\cos\frac{\pi\beta}{2} \Gamma(\beta+1)}{(K|y_{\rm end}|)^\beta}\left[\sum_j k_j^3 +\frac{\beta -1}{2} \sum_{i\neq j}k_i k_j^2 - 2(\beta+3) k_1 k_2 k_3 \right. \right.\\ \nonumber
&-& \left. \frac{2(\beta+1)}{K^2} \sum_{i\neq j} k_i^2 k_j^3 + \frac{3}{2 K k_1 k_2 k_3} \left(\sum_{i \neq j} k_i^2 k_j^5 - \sum_{i \neq j} k_i^3 k_j^4 \right)\right] \\ \nonumber
&+& 3 \left(\frac{\cos\frac{\pi\beta}{2} \Gamma(\beta)}{(K|y_{\rm end}|)^\beta} -\frac{1}{\beta}\right)
\left[-9 k_1 k_2 k_3 - 2 \sum_{i \neq j} k_i k_j^2 + \right. \\ 
&+& \left. \frac{1}{k_1 k_2 k_3} \left(\frac{1}{2} \sum_{i\neq j}k_i k_j^5
 -  \sum_{i< j}k_i^3 k_j^3\right)+   \frac{1}{2 k_1^2 k_2^2 k_3^2} \left(\sum_{i\neq j}k_i^3 k_j^6 - \sum_{i\neq j}k_i^4 k_j^5 \right)\right]\nonumber \\
&+& 3 K \left(\frac{\cos\frac{\pi\beta}{2} \Gamma(\beta-1)}{(K|y_{\rm end}|)^\beta} +\frac{1}{\beta}\right)\left[- 2 \sum_j k_j^2 - 9 \sum_{i< j}k_i k_j + \frac{1}{k_1^2 k_2^2 k_3^2}\left( \frac{1}{2} \sum_{i\neq j}k_i^2 k_j^6 - \sum_{i< j}k_i^4 k_j^4 \right)\right]\nonumber \\
&-& \left. 27\left[\frac{\cos\frac{\pi\beta}{2}}{(K|y_{\rm end}|)^\beta}\frac{\Gamma(\beta)}{\beta-2} + \frac{\beta-1}{\beta(\beta-2)}-\frac{1}{3}\right] \frac{K^3}{\beta-3} \right\}\, , \nonumber 
\eea
\bea
\nonumber
{\cal A}^{(\bf II)}_{\epsilon^2} &=& \frac{\epsilon^2}{32\, c_{s\ \rm end}^2 }\left\{\frac{\cos\frac{\pi\beta}{2} \Gamma(\beta+1)}{(K|y_{\rm end}|)^\beta}\left[(\beta+4) \left(\sum_j k_j^3 - \sum_{i\neq j}k_i k_j^2 + 2 k_1 k_2 k_3\right)\right. \right.\\ \nonumber
&-& \left.12 k_1 k_2 k_3 - \frac{3}{K} \sum_{i\neq j} k_i k_j^3 + \frac{3}{K k_1 k_2 k_3}(\sum_{i \neq j} k_i^2 k_j^5 - 2 \sum_{i \neq j} k_i^3 k_j^4 + \sum_{i \neq j} k_i k_j^6)\right] \\ \nonumber
&+& \left(\frac{\cos\frac{\pi\beta}{2} \Gamma(\beta)}{(K|y_{\rm end}|)^\beta} -\frac{1}{\beta}\right)
\left[-54 k_1 k_2 k_3 - 15 \sum_{i \neq j} k_i k_j^2 - 6 \sum_j k_j^3 + \frac{3}{k_1 k_2 k_3} \left(3 \sum_j k_j^6 \right. \right. \\
&-&\left. \left.  3 \sum_{i\neq j}k_i^2 k_j^4 + 2 \sum_{i\neq j}k_i k_j^5 - 4 \sum_{i< j}k_i^3 k_j^3\right)+   \frac{3}{k_1^2 k_2^2 k_3^2} \left(\sum_{i\neq j}k_i^3 k_j^6 
+ \sum_{i\neq j}k_i^2 k_j^7 - 2 \sum_{i\neq j}k_i^4 k_j^5\right)\right]\nonumber \\
&+& 3 K \left(\frac{\cos\frac{\pi\beta}{2} \Gamma(\beta-1)}{(K|y_{\rm end}|)^\beta} +\frac{1}{\beta}\right)\left[- 6 \sum_j k_j^2 - 18 \sum_{i< j}k_i k_j - \frac{3}{k_1 k_2 k_3}\left( \sum_{i\neq j}k_i^2 k_j^3 +\sum_{i\neq j}k_i k_j^4 \right. \right. \nonumber \\
&-&\left.\left. \sum_j k_j^5\right) + \frac{1}{k_1^2 k_2^2 k_3^2}\left( 2 \sum_{i\neq j}k_i^2 k_j^6 -4 \sum_{i< j}k_i^4 k_j^4 + 3 \sum_{i\neq j}k_i k_j^7 - 3 \sum_{i\neq j}k_i^3 k_j^5\right)\right] \nonumber \\
&-& \left. 9\left[\frac{\cos\frac{\pi\beta}{2}}{(K|y_{\rm end}|)^\beta}\frac{\Gamma(\beta)}{\beta-2} + \frac{\beta-1}{\beta(\beta-2)}-\frac{1}{3}\right] \frac{K^3}{\beta-3} \left[6 - \frac{1}{k_1^2 k_2^2 k_3^2} \left(\sum_j k_j^6 -  \sum_{i\neq j}k_i^2 k_j^4\right) \right] \right\} \nonumber \ .
\eea
Finally, there is a term associated with the field redefinition of $\zeta$:
\bea
{\cal A}^{(\bf II)}_{\rm redef} &=&   -\frac{3(1+\epsilon)}{10\, c_{s\;{\rm end}}^2}\left[\sum_j k_j^3 \right]\, .
\label{3ampII}
\eea

We will use two different measures for the amount of non-Gaussianity:
\beq
f_{\rm NL} \equiv \frac{40}{3} \frac{1}{K^3} \ \lim_{k_1 \to 0} {\cal A}\, ,
\eeq
for non-Gaussianity in the squeezed limit, and
\beq
\widehat f_{\rm NL} \equiv 30\, \frac{1}{K^3}\  {\cal A}_{k_1=k_2=k_3}\, ,
\eeq
for non-Gaussianity of general shapes.

\subsection{Squeezed Limit}

We consider the squeezed limit ($k_1 \to 0$, $k_2 \to k_3 = \frac{K}{2}$) of the above bispectra.
For the first interaction we find:
\bea
\nonumber
f_{\rm NL}^{\dot\zeta^3} &=& \frac{4}{3} \frac{(1+\epsilon)}{c_{s\;{\rm end}}^{2}} \cdot \left(f_X^{(0)} - 1 \right) \cdot \left\{\frac{\cos\frac{\pi\beta}{2}}{(K|y_{\rm end}|)^\beta} \frac{\Gamma(\beta+1)}{4} \left[ \frac{3}{4} + \frac{9}{\beta-1} \right]  \right. \\ \nonumber
&& \hspace{1cm}+  \left. \frac{9}{(\beta-2)(\beta-3)}\left[\frac{\cos\frac{\pi\beta}{2} }{(K|y_{\rm end}|)^\beta} \, \Gamma(\beta)\, \frac{\beta^2-2\beta+3}{\beta-1} + \frac{3}{\beta}\right]  \right\}
 \,.\nonumber 
\eea
In the limit $\beta \to 0$ this gives
\bea
\lim_{\beta \to 0}\ f_{\rm NL}^{\dot \zeta^3} &=& \frac{f_X^{(0)}-1}{c_{s\, \rm end}^2} \Bigl( 24 \cdot \log (K y_{\rm end}) -19+24 \gamma_E \Bigr) \ \  + \ \  {\cal O}(\beta)\ .
\eea
For generic values of $f_X^{(0)} \sim {\cal O}(1)$ this gives a relatively large scale-dependent non-Gaussianity.
However, for the DBI action $f_{X}^{(0)} =1$ and the leading term doesn't contribute.
The next interaction has the following squeezed limit
\bea
f_{\rm NL}^{\zeta (\dot \zeta)^2} &=& \frac{5(\epsilon-3)}{12\, c_{s \ \rm end}^2} \Bigl\{ \frac{\cos \frac{\pi \beta}{2}}{(K|y_{\rm end}|)^\beta} \left[ \Gamma(\beta+1) + 12\, \Gamma(\beta) \frac{\beta^2 - 5 \beta + 12}{(\beta-2)(\beta-3)}+ 78\, \Gamma(\beta-1)\right] \nonumber\\
&&\hspace{1.5cm} +\ \frac{6}{\beta}\, \frac{2\beta^2 -10 \beta-15}{(\beta-2)(\beta-3)} \Bigr\} + \frac{5}{2}\ ,
\eea
which for small $\beta$ becomes
\bea
\lim_{\beta \to 0}\  f_{\rm NL}^{\zeta (\dot \zeta)^2} &=& \left(\frac{5}{2} - \frac{23}{c_{s\, \rm end}^2} \right) \ \ + \ \ {\cal O}(\beta)\ .
\eea
Next, we have
\bea
f_{\rm NL}^{\zeta(\vec{\nabla}\zeta)^2} &=& \frac{13 \epsilon-7}{24 c_{s\ \rm end}^2} \frac{\cos \frac{\pi \beta}{2}}{(K y_{\rm end})^\beta} \Gamma(\beta+1) \frac{\beta + 3}{\beta-1} \ ,
\eea
which for small $\beta$ is
\bea
\lim_{\beta \to 0}\  f_{\rm NL}^{\zeta(\vec{\nabla}\zeta)^2} &=& - \frac{4}{c_{s\, \rm end}^2} \ \ + \ \ {\cal O} (\beta) \ .
\eea
For the two non-local interactions we find
\bea
f_{\rm NL}^{\dot \zeta \partial \zeta \partial \chi}   &=& \frac{30 \epsilon}{c_{s\ \rm end}^2} \left[ \frac{1}{8}  \left(\frac{\cos\frac{\pi\beta}{2} \Gamma(\beta)}{(K|y_{\rm end}|)^\beta} -\frac{1}{\beta}\right) + \left(\frac{\cos\frac{\pi\beta}{2} \Gamma(\beta-1)}{(K|y_{\rm end}|)^\beta} +\frac{1}{\beta}\right) \right. \nonumber \\
&& \hspace{2cm} \left. + \frac{3}{\beta-3} \left(\frac{\cos\frac{\pi\beta}{2}}{(K|y_{\rm end}|)^\beta}\frac{\Gamma(\beta)}{\beta-2} + \frac{\beta-1}{\beta(\beta-2)}-\frac{1}{3}\right) \right] \ , \\
f_{\rm NL}^{\epsilon^2}  &=& - \frac{\epsilon}{3} \cdot f_{\rm NL}^{\dot \zeta \partial \zeta \partial \chi}   \ .\label{equ:epsSqr} 
\eea
It is interesting to consider the limit $\beta \to 0$ ($\epsilon \to 3$). In this case, we get, for example,
\bea
\lim_{\beta \to 0}\ f_{\rm NL}^{\dot \zeta \partial \zeta \partial \chi}   &=& \frac{15}{4 c_{s\ \rm end}^2} (-4 + 9 \gamma_E) + \frac{135}{4 c_{s\ \rm end}^2} \log(K y_{\rm end}) \ + \ {\cal O}(\beta)\ .
\eea
This looks like a dangerously large amount of non-Gaussianity arising from the superhorizon evolution of $\zeta$.
However, from (\ref{equ:epsSqr}) we get an equal contribution with opposite sign for $\beta=0$, so that the sum is proportional to $\beta$,
\bea
\lim_{\beta \to 0} \ ( f_{\rm NL}^{\dot \zeta \partial \zeta \partial \chi} + f_{\rm NL}^{\epsilon^2} ) &=&  \frac{\beta}{c_{s\, \rm end}^2} \Bigl[ (-4 + 9 \gamma_E) + 9 \log(K y_{\rm end}) \Bigr] \ + \ {\cal O}(\beta^2)\ .
\eea
For $\beta \ll (\log (K y_{\rm end}))^{-1}$ and $c_{s \, \rm end} \sim 1$ the non-local interactions are therefore less dangerous than they appear at first sight.\footnote{This cancellation didn't happen in \cite{Justin} due to algebraic errors in their bispectrum computation.}
However, it is worth pointing out that this cancellation only holds in the squeezed limit.
For the equilateral configuration, for example, we find
\bea
\lim_{\beta \to 0} \ ( \widehat f_{\rm NL}^{\ \dot \zeta \partial \zeta \partial \chi} + \widehat f_{\rm NL}^{\, \epsilon^2} ) &=&  - \frac{1}{c_{s\, \rm end}^2} \frac{375}{32} \log(K y_{\rm end}) + \frac{2.3}{c_{s\, \rm end}^2}  \ + \ {\cal O}(\beta)\ .
\eea
Finally, from the field redefinition we get
\bea
f_{\rm NL}^{\rm redef} &=& - \frac{1+\epsilon}{c_{s \, \rm end}^2} \ \ \to\ \ - \frac{4}{c_{s \, \rm end}^2} \left[1 + \frac{\beta}{5} \right] \, .
\eea
In \S\ref{sec:classical} we present results for $f_{\rm NL}$ and $\widehat f_{\rm NL}$ as a function of $\beta$.

\vfil

\end{document}